\documentclass[%
 reprint,
nofootinbib,
nobibnotes,
 amsmath,amssymb,
 aps,prd,
floatfix,
]{revtex4-2}

\usepackage{graphicx}
\usepackage{dcolumn}
\usepackage{bm}
\usepackage{hyperref}
\usepackage[mathlines]{lineno}
\usepackage{graphicx}
\usepackage{dcolumn}
\usepackage{bm}
\usepackage{hyperref}
\usepackage{color}
\usepackage{natbib}
\usepackage{CJK}
\usepackage[normalem]{ulem}
\usepackage{comment}

\usepackage[english]{babel}
\usepackage{xspace}
\usepackage{ulem}

\hypersetup{
    colorlinks=true,
    linkcolor=magenta,
    citecolor=blue,
}

\newcommand{\apjl}{Astrophys.\ J.\ }

\newcommand{\aap}{Astron.\ Astrophys.\ }

\newcommand{\physrep}{Phys.\ Rep.\ }
\newcommand{\mnras}{Mon.\ Not.\ R.\ Astron.\  Soc.\ }

\newcommand{\nphysa}{Nucl.\ Phys.\ A.}

\newcommand{\nar}{New Astron.\ Rev.\ }

\begin{document}

\preprint{APS/123-QED}

\title[Cocoon shock breakout emission from binary neutron star mergers]{Cocoon shock breakout emission from binary neutron star mergers}

\author{Eduardo M. Guti\'errez$^{1,2}$, Mukul Bhattacharya$^{2,3,4}$, David Radice$^{1,2,3}$, Kohta Murase$^{2,3}$, and Sebastiano Bernuzzi$^5$}
\affiliation{%
$^1$Institute for Gravitation and the Cosmos, The Pennsylvania State University, University Park, Pennsylvania, 16802, USA \\
$^2$Department of Physics, The Pennsylvania State University, University Park, Pennsylvania, 16802, USA \\
$^3$Department of Astronomy and Astrophysics, The Pennsylvania State University, University Park, Pennsylvania, 16802, USA \\
$^4$Department of Physics, Wisconsin IceCube Particle Astrophysics Center, University of Wisconsin, Madison, Wisconsin, 53703, USA \\
$^5$Theoretisch-Physikalisches Institut, Friedrich-Schiller-Universit¨at Jena, 07743, Jena, Germany
}

\date{\today}

\begin{abstract}
Shock breakout emission is likely the first electromagnetic signal from a wide variety of astrophysical explosive phenomena, including supernovae and neutron star (NS) mergers; as exemplified by GRB 170817A, this signal can be the dominant component in low-luminosity short $\gamma$-ray bursts.
In this work, we investigate the cocoon shock breakout emission in NS mergers and how its signal depends on the outermost layers of the ejecta profile, which we derive from general-relativistic radiation hydrodynamic simulations.
To explore the influence of the outermost layers of the ejecta on the breakout emission, we model the ejecta profile as either having a sharp cutoff or an extended smooth tail.
We find that the shock breakout emission is strongly influenced by the shape of the outermost layers of the ejecta, with breakouts from extended density profiles yielding emission consistent with the observed properties of GRB 170817A; on the contrary, breakouts from ejecta with a sharp cutoff tend to overestimate the radiated energy.
Using a Bayesian analysis, we estimate the best-fit parameters for the central engine, considering both accreting black hole (BH) and magnetized NS scenarios.
Our findings indicate a slight preference for scenarios in which the remnant suffers an early collapse to a BH.
Our work probes the nature of NS mergers and highlights the importance of carefully treating the shape of the ejecta outermost layers in modeling early electromagnetic counterparts from these events.

\end{abstract}

\maketitle

\section{Introduction} \label{sec:intro}

	The event GW170817 represented not only the first detection of gravitational waves (GWs) from the merger of two neutron stars (NSs) but also the first multimessenger event with GWs and electromagnetic radiation \cite{LIGO_GW170817, LIGO_MM_2017}.
	On August 17, 2017, the LIGO Scientific Collaboration and the Virgo Collaboration simultaneously detected a GW signal from the merger of two compact objects with total mass $2.82^{+0.47}_{-0.09} M_\odot$, consistent with the components being two NSs with masses $m_1 \in (1.36, 2.26) M_\odot$ and $m_2 \in (0.86, 1.36) M_\odot$.
	Approximately $2$ s after this detection, a short $\gamma$-ray burst (sGRB) was independently detected by the \emph{Fermi} Gamma-Ray Burst Monitor telescope and the SPI-ACS instrument onboard the International Gamma-Ray Burst Astrophysics Laboratory \cite{LIGO_GRB_2017}; the emission came from a region consistent with the localization derived from the GW signal.
	Hours later, a thermal optical counterpart, AT2017gfo, was detected, which allowed the identification of the host galaxy, NGC 4993, located at a distance of ${\approx} 40 ~{\rm Mpc}$ \cite{LIGO_MM_2017, Coulter_etal2017}.
	The optical emission was consistent with a radioactive-decay-powered kilonova originating from the NS merger \cite{Drout_etal2017, Metzger2019}; this component also produced detectable radiation in the ultraviolet and infrared bands.
	Later, emission potentially associated with the afterglow was detected in x-rays \cite{Troja_etal2017} and radio \cite{Hallinan_etal2017}.
	Late-time very long baseline interferometry radio observations demonstrated the presence of a relativistic jet \cite{Mooley_etal2018Natur.561..355M, Balasubramanian_etal2021}.
	
	The $\gamma$-ray detection event, hereafter referred to as GRB 170817A, confirmed the long-standing hypothesis that sGRBs can originate from NS/NS black hole (BH) mergers \cite{Blinnikov_etal1984, Paczynski1986, Eichler_etal1989, Narayan_etal1992}.
	However, the luminosity of this sGRB was much lower than that of a regular event.
 The main peak of the GRB, which arrived $t_{\rm GRB}=(1.73 \pm 0.05)~{\rm s}$ after the arrival of the GWs, had a fluence of $(2.8 \pm 0.2)\times 10^{-7}~{\rm erg}~{\rm cm}^{-2}$ \cite{Goldstein_etal2017_848L..14G}.
	A simple estimate using the inferred distance to the source gives an isotropic equivalent energy of $E_{\rm iso} = (5.1 \pm 1) \times 10^{46}~{\rm erg}$, which is ${\sim} 4$ orders of magnitude lower than the average value for the detected sGRB population (see, e.g., Ref. \cite{Nakar2020}).
    The best fit for the spectrum of the primary $\gamma$-ray pulse is a power law with an exponential cutoff with a spectral index of $\alpha=-0.62 \pm 0.40$ and peak energy of $E_{\rm pk,GRB} = (185 \pm 62)~{\rm keV}$  \cite{Goldstein_etal2017_848L..14G}.
	Because of its unique properties among the known sGRBs, this event is sometimes considered to represent a different type of GRB: a low-luminosity short GRB.
	Unfortunately, GRB 170817A is to date the only sGRB event with these characteristics occurring close enough to be detectable.
 
	Several scenarios for the origin of such weak $\gamma$-ray emission have been proposed.
 These include an intrinsically weak relativistic jet whose emission was observed on axis, a regular powerful relativistic jet whose emission was observed off axis, and a shock breakout emission produced by a structured jet or a cocoon propagating through the NS ejecta \cite{Alexander_etal2017ApJ...848L..21A, Margutti_etal2017ApJ...848L..20M, Troja_etal2017, Haggard_etal2017ApJ...848L..25H, Kasliwal_etal2017}.
	The third scenario was first proposed in Ref. \cite{Kasliwal_etal2017} and later investigated by several other authors \cite{Mooley_etal2018Natur.554..207M, Bromberg_etal2018MNRAS.475.2971B, Gottlieb_etal2018b, Nakar_etal2018, BeloborodovLundman2020, Nakar2020, Fraija_etal2019}, and it has shown to provide a natural explanation for most properties of the observed $\gamma$-ray signal.
	The intrinsically weak jet scenario faces challenges related to the jet's ability to break out of the ejecta successfully, whereas the off axis jet scenario requires fine-tuning of viewing angle and jet structure parameters to account for the $\gamma$-ray emission \cite{Kasliwal_etal2017}.
    Late-time radio observations extending to $\sim 300$ days after the merger provided evidence consistent with an off axis jet interpretation for the afterglow \cite{Mooley_etal2018ApJ...868L..11M}.
    These findings suggest that the merger may have given rise to a jet+cocoon system that successfully broke out of the ejecta; the wide-angle cocoon breakout likely produced the $\gamma$-ray emission, while the jet produced the late-time afterglow emission.
    	
    An important aspect influencing shock breakout emission is how the jet propagates through the ambient medium.
    Several authors have studied the propagation of relativistic jets through a dense ambient medium \cite{RamirezRuiz_etal2002, Matzner2003, Morsony_etal2007, Bromberg_etal2011}.
    Later, these treatments were extended for the case of an expanding medium \cite{Salafia_etal2020, HamidaniIoka2021, Gottlieb_etal2023} and magnetized jets \cite{LevinsonBegelman2013, Bromberg_etal2014,Bhattacharyaetal_2023,Bhattacharyaetal_2024}.
	Our main aim in this work is to develop a model to calculate the emission from shocks driven by the cocoon breaking out of the NS merger ejecta for a wide range of central engine properties and ejecta profiles.
    To this end, we apply a similar approach to these works, though with small modifications to treat more arbitrary ejecta profiles and allow time-dependent engine luminosities.
	We obtain the NS merger ejecta properties from a set of general-relativistic neutrino-radiation hydrodynamics long-term simulations, with different mass ratios and/or nuclear equation of state (EOS), though all consistent with GW170817.
	
    The outflows derived from these simulations demonstrate the presence of a fast component with speed $v>0.5c$, the so-called fast ejecta, which naturally constitutes the outermost regions of the outflow.
    However, the specific shape of these outer layers is largely unconstrained once the outflow expands to large distances since simulations are affected by resolution, floor effects, and finite domains.
	One possibility is that the ejecta presents a rather sharp cutoff at the outer boundary determined by those parcels of plasma that achieve the highest velocity shortly after the merger.
    Another possibility is that because of the intrinsic thermal velocity dispersion of the ejecta matter or due to the interaction between different layers of the outflow, the ejecta develops a smooth extended tail without a clear cutoff.
        The specific properties of the outermost layers of the ejecta may have a large influence on the amount of matter involved during the breakout emission and hence on the emitted signal \cite{BeloborodovLundman2020, Nakar2020}.
	To investigate this effect in detail, we derive radial ejecta profiles from numerical relativity simulations and consider two cases: (i) the ejecta profile has a ``sharp cutoff'' at $r\sim r_{\rm max}(t)$ and (ii) the ejecta has an ``extended tail'' originated in the thermal dispersion of matter in the outermost layers.
	Once we have determined the ejecta profiles, we model the central engine that launches the relativistic jet based on the nature of the remnant left behind after the merger; namely, we distinguish the scenarios where the remnant is an accreting black hole or a highly magnetized NS.
	We investigate the propagation of the jet through the ejecta, the cocoon evolution before and after the jet breaks out of the ejecta, and finally, we estimate the electromagnetic emission released when the forward shock driven by the cocoon itself breaks out of the ejecta.
	We compare the observables derived from our model for a large set of engine parameters with observables from GRB 170817A, assuming that this event indeed originated from a cocoon shock breakout.
	Finally, we infer the most likely engine parameters and discriminate among the various simulated ejecta profiles and the two approaches (sharp cutoff or extended tail) considered to model their outer layers.

    We find that ejecta profiles with an extended tail produce electromagnetic signals at break out consistent with that of GRB 170817A, whereas those with a sharp cutoff at the outermost boundary systematically overestimate the radiated energy.
    Regarding merger remnant and jet engines, BH engines are marginally favored over magnetar engines, though we can only rule out one of the three models that produce a long-lived NS remnant.
 
	The paper is structured as follows.
	In Sec.~\ref{sec:physical_scenario}, we present the details of our physical model.
	We summarize the main characteristics of the numerical relativity (NR) simulations considered to model the ejecta profiles, and discuss the parametrizations used for the engine and jet properties.
    We derive the equations for the jet and cocoon evolution as they propagate in the ejecta until the time when the jet finally breaks out.
	Then, we describe our treatment for the cocoon evolution once the jet has broken out of the ejecta but most of the cocoon is still trapped.
	In Sec.~\ref{sec:results}, we analyze the results from each of our simulations, in particular, the jet and cocoon evolution, with a special focus on the properties of these components at the time of breakout.
	Then, in Sec.~\ref{sec:shockbreakout}, we discuss the cocoon shock breakout emission and calculate the bolometric light curves associated with it.
	In Sec.~\ref{sec:GW170817}, we compare the main observables derived from our calculations with those of GRB 170817A. 
	We perform a Bayesian analysis to compare the likelihood of the five NR simulations in order to estimate the most likely engine parameters.
	In Sec.~\ref{sec:discussion}, we discuss the most important points that we have derived from our analysis and compare them with those in the literature.
	Finally, we present our conclusions and discuss possible future work in Sec.~\ref{sec:conclusions}.

\section{Physical Scenario} \label{sec:physical_scenario}

\begin{figure*}
    \centering
    \includegraphics[width=0.99\linewidth]{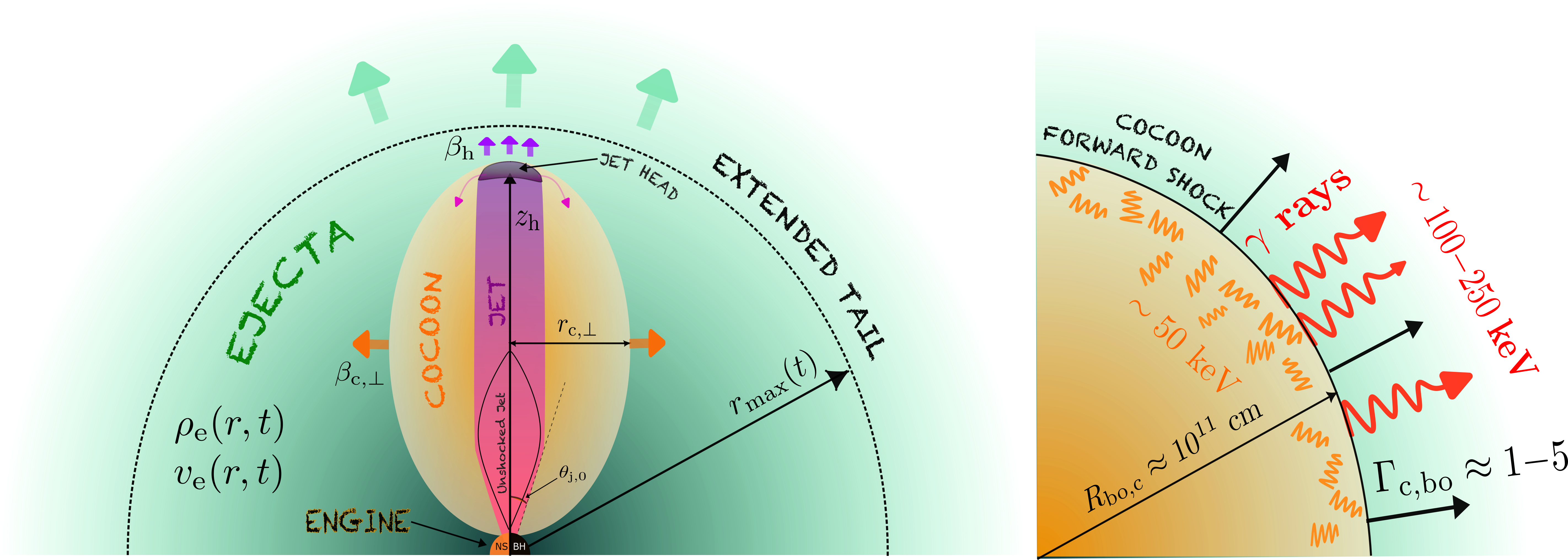}
    \caption{
    Schematic representation of the various physical components of our model. Left: the jet+cocoon system propagating through the ejecta of the NS merger.
    We show the jet, with its unshocked region, the jet’s head, the cocoon, and the ejecta, for which we identify the boundary for the sharp cutoff [at $r_{\rm max}(t)$] scenario as well as the extended tail.
The head expands through the ejecta with speed $\beta_{\rm h}$ in the vertical direction, whereas the cocoon also expands laterally with speed $\beta_{\rm c,\perp}$.
We indicate the position of the head $z_{\rm h}$ and the lateral radius of the cocoon $r_{\rm c, \perp}$, as well as the rest-mass density and velocity profile of the ejecta: $\rho_{\rm ej}(r,t)$, $v_{\rm ej}(r,t)$, respectively.
Right: a closeup of the cocoon and its forward shock at the time of breakout $t_{\rm bo,c}$ at a radius of $R_{\rm bo,c}\approx 10^{11}~{\rm cm}$.
Pairs regulate the downstream temperature to be ${\sim} 50~{\rm keV}$; when these photons are released, they are observed with an energy ${\sim} 100{-}250~{\rm keV}$ since the Lorentz factor of the breakout layer is $\Gamma_{\rm c,bo}\approx 1{-}5$.}
    \label{fig:cartoon}
\end{figure*}

	The physical scenario considered here and the various components in our model are shown schematically on the left panel of Fig. \ref{fig:cartoon}.
	We assume that two NSs with masses $M_1$ and $M_2$ merge at time $t=0$, leading to the ejection of matter in all directions.
	The resulting mass outflow forms a dense, mildly relativistic expanding medium.
	The merger leaves behind a compact remnant, whose nature depends mainly on the total mass of the binary, $M=M_1 + M_2$, its mass ratio, $q=M_1/M_2$, and the microphysical properties of matter encoded mainly in the unknown nuclear EOS.
	The merger outcome and the nature of the remnant can be roughly separated into three cases:
\begin{enumerate}
	\item[(1)] Early collapse: The remnant is too massive ($M > M_{\rm TOV}$, where $M_{\rm TOV}$ is the maximum mass for a Tolman--Oppenheimer--Volkoff star for a given EOS) and collapses into a BH on a timescale of $t \lesssim 15 ~ {\rm ms}$.
	Because of the short collapse time, most of the matter gets trapped within the event horizon.
 The remaining matter may form a light accretion disk around the BH.
	\item[(2)] Delayed collapse: The remnant has a mass $M \gtrsim M_{\rm TOV}$ and survives for some time as a differentially rotating hypermassive neutron star (HMNS).
    This occurs until the HMNS loses sufficient rotational energy through GWs and winds and collapses into a BH at some time between $t \sim {\rm tens~of~milliseconds}$ and $t \sim {\rm few ~ seconds}$.
	\item[(3)] Long-lived NS: The NS remnant has a mass $M \lesssim M_{\rm TOV}$ and does not collapse into a BH during the timescales of interest ($t < 10~{\rm s}$).
\end{enumerate}
	
\subsection{Merger ejecta}

    On timescales $t \lesssim 10 ~{\rm ms}$ after the merger, the outflows are dominated by the ``dynamical ejecta'', namely, the ejecta driven by dynamical forces as a direct consequence of the collision \cite{Bauswein_etal2013, Hotokezaka_etal2013, Wanajo_etal2014ApJ...789L..39W, Foucart_etal2016PhRvD..93d4019F, KastaunGaleazzi2015PhRvD..91f4027K, Palenzuela_etal2015PhRvD..92d4045P, Sekiguchi_etal2015PhRvD..91f4059S, Radice_etal2018, Karakas_etal2024arXiv240513687K}.
    On larger timescales ($t > 10 ~ {\rm ms}$), the properties of the ejecta depend largely on the outcome of the remnant formed after the NS collision \cite{Dessart_etal2009ApJ...690.1681D, MetzgerFernandez2014MNRAS.441.3444M, Perego_etal2014MNRAS.443.3134P, Fernandez_etal2015MNRAS.446..750F, Just_etal2015MNRAS.448..541J, Martin_etal2015ApJ...813....2M, Nedora_etal2019ApJ...886L..30N, Shibata_etal2019PhRvD.100b3015S,
    Nedora_etal2021ApJ...906...98N, Just_etal2023ApJ...951L..12J, RadiceBernuzzi2023}.
	If the total binary mass is large enough ($\gtrsim 2.8 M_\odot$), then the NS collision leads to rapid collapse to a BH, and, as a result, the dynamical mass ejection significantly drops~\cite{Shibata_etal2005PhRvD..71h4021S, Radice_etal2018, Radice_etal2020, CombiSiegel2023a, Kiuchi_etal2023PhRvL.131a1401K}.
    
\begin{itemize}
	\item[(i)] Dynamical ejecta:
	The dynamical ejecta is driven both by gravitational forces and by shock heating at the collision interface between the two NSs.
	The shock-driven ejecta generates a largely isotropic fast component of the outflow, whose asymptotic velocities can be mildly relativistic: $v_{\rm fast} \sim 0.5{-}0.8c$.
	In contrast, the tidal component of the dynamical ejecta is predominantly launched close to the equatorial plane, with slower speeds ($v\sim 0.1{-}0.4c$) and a low electron fraction ($Y_{\rm e} \sim 0.1$).
	This neutron-rich component of the ejecta is believed to be mainly responsible for the radioactive-decay-powered kilonova at later times.
	
	The amount of dynamical ejecta is strongly influenced by the binary mass ratio $q=M_1/M_2$.
	For large mass ratios, the lighter star will be tidally disrupted before the merger itself, leading to a much weaker collision and thus producing a lower amount of shock-heated ejecta.
	Conversely, the tidally ejected outflow will be enhanced.
	
	\item[(ii)] Secular ejecta: Part of the material disrupted from the stars remains gravitationally bound and forms a thick accretion disk around the central remnant, either an NS or a BH.
	The dynamics of these dense disks is mediated by neutrinos, and so they are called neutrino-dominated accretion disks \cite{Popham_etal1999, Narayan_etal2001, Liu_etal2017}.
	The neutrinos heat matter to drive a baryonic wind from the surface of the accretion disk.
	If the central remnant has not collapsed to a BH, the surface of the central NS can also be a strong source of neutrinos and contribute to the outflow \cite{Dessart_etal2009ApJ...690.1681D, Perego_etal2014MNRAS.443.3134P, Nedora_etal2021ApJ...906...98N, Just_etal2023ApJ...951L..12J}.
    This component may dominate the ejecta in the polar region, even though the total ejecta mass is dominated by the wind launched by spiral density waves in the remnant, which transport angular momentum and energy from the core to the outer layers, driving mass ejection.
	The neutrino-driven wind can be sustained up to ${\sim} 10~{\rm s}$ and typically has masses of $\sim 10^{-3}{-}10^{-2}\,M_{\odot}$, a higher electron fraction $Y_{\rm e} \gtrsim 0.2$ due to intense neutrino irradiation, and velocities in the range $v \sim 0.1{-}0.2c$.

\end{itemize}	

	The different components of the ejected outflow expand and mix, forming a dense stratified expanding medium.
	For this work, we characterize this expanding medium by means of two functions of the radial coordinate and the time: the rest-mass density profile $\rho_{\rm ej}(r,t)$ and the velocity profile $v_{\rm ej}(r,t)$.

\subsubsection{Numerical relativity simulations} \label{sssec:simulation}

    To model the expanding ejecta profile, we consider a set of five long-term NR simulations performed with the code {\tt THC\_M1} \cite{Radice_etal2014a, Radice_etal2014b, Radice_etal2015, Radice_etal2022}, which represent different NS merger scenarios.
    {\tt THC\_M1} is a general-relativistic hydrodynamics code based on the {\tt Einstein Toolkit} \cite{Loffler_etal2012} coupled with a moment-based energy-integrated scheme to treat the transport of neutrinos \cite{Radice_etal2022}.
    This scheme is more accurate than the most popular M0 + leakage schemes, since it correctly handles neutrino trapping in relativistically moving media, such as rotating NS remnants.
    
    The five simulations we consider here differ either in the nuclear EOS or in the binary mass ratio, and produce different outcomes: some of them give rise to an early collapse to a BH whereas others give rise to a long-lived remnant.
    All simulations but SLy\_M145-125 make use of the general-relativistic large-eddy simulation formalism \cite{Radice2017, Radice2020} to account for the angular momentum transport caused by magnetohydrodynamic turbulence.
    For more details, see \cite{RadiceBernuzzi2023}.
    
    In the following, we list the main properties of each simulated merger.
    Note that the isotropic equivalent masses\footnote{The isotropic equivalent mass is defined as $M_{\rm iso}=4\pi \int dr r^2 \rho_{\rm ej}(r)$, where $\rho_{\rm ej}(r)$ is the ejecta rest-mass density profile in the polar region.} listed below are measured by integrating the mass-loss rate over the duration of each simulation, but this quantity can keep increasing at later times.
	
\begin{itemize}
	\item[(i)] DD2\_M135-135: Equal-mass binary with the DD2 EOS \cite{Hempel_SchaffnerBielich2010, Typel_etal2010}.
 This is a stiff EOS with a large allowed maximum TOV mass: $M_{\rm TOV}^{\rm max}$.
	The simulation ran for ${\sim} 98~{\rm ms}$, over which the remnant did not collapse.
	The ejecta has a fast component with a maximum asymptotic speed of $v\simeq 0.58c$, followed by a steady secular ejecta component driven by neutrino irradiation from the NS remnant and the accretion disk.
	The total isotropic equivalent mass in the polar region is ${\approx} 0.003 M_\odot$.
 This simulation was analyzed in detail in Refs.~\cite{RadiceBernuzzi2023, RadiceBernuzzi2024}.
	\item[(ii)] DD2\_M180-108: Same EOS as in the previous case, but here the NSs have a large mass ratio $q \simeq 1.8$.
	This simulation ran for ${\sim} 122~{\rm ms}$ without producing a collapse.
	The ejecta for this case has the largest total isotropic equivalent mass among the five simulations we considered, ${\approx} 0.01 M_\odot$.
	The fastest ejecta also reaches a significantly larger speed of $v\sim 0.76c$.
	\item[(iii)] BLh\_M1146-1635: This simulation considered a binary mass ratio $q\approx 1.4$ and the BLh EOS \cite{BombaciLogoteta2018}.
    It ran for ${\sim} 100~{\rm ms}$, over which the remnant did not collapse.
	The ejecta has a maximum asymptotic velocity of $v\sim 0.66c$ and a total equivalent mass of ${\approx} 0.009 M_\odot$.
	\item[(iv)] SFHo\_M135-135: The binary in this simulation has a mass ratio $q=1$ and uses the soft SFHo EOS \cite{Steiner_etal2013}.
 The merger produced a short-lived remnant that collapses to a BH at $t\sim 15$ ms after the merger.
	The simulation ran for ${\sim}57$ ms.
	The ejecta in the polar region has a total equivalent mass of ${\approx} 0.002 M_\odot$, the smallest among the five simulations, and its fast component reaches a maximum asymptotic velocity of $v\sim 0.78c$.
 \item[(v)] SLy\_M145-125: The binary in this simulation has a mass ratio of $q\approx 1.2$, and uses the soft SLy4 EOS \cite{Chabanat_etal1998, Schneider_etal2017}.
 It also produced a short-lived remnant that collapsed to a BH at $t\approx 11$ ms after the merger.
	The mass-loss rate rapidly decreases after BH formation, and the simulation ran for ${\sim} 35~{\rm ms}$.
	The fast ejecta reaches a velocity of $v\sim 0.73c$ and the ejecta has a total equivalent mass of ${\approx} 0.007 M_\odot$.
 This simulation was analyzed in detail in Refs.~\cite{Espino_etal2024a, Espino_etal2024b}.
\end{itemize}
    The simulations DD2\_M180-108, BLh\_M1146-1635, and SFHo\_M135-135 will be analyzed in full detail in an upcoming paper (Bernuzzi {\it et al.} \cite{Bernuzzi_inprep2025}).
    
    For each simulation, the azimuthally averaged outflow properties are recorded at each polar angle over a sphere placed at a radius $R_0 \simeq 442~{\rm km}$\footnote{This radius (which corresponds to $300$ in code units) is chosen so that most of the captured ejecta is ballistic, while at the same time its density is still high enough to be unaffected by the numerical atmosphere.}.
    These properties include the rest-mass density $\rho$, the outflow radial velocity $v$, the temperature $T$, the specific entropy $s$, the proton fraction $Y_{\rm e}$, and the Bernoulli parameter $\mathcal{B}=-hu_t$.
    The latter coincides with the asymptotic Lorentz factor $\Gamma_\infty$.
    We assume that for the distances of interest in our work, $r \gg R_0$, the ejecta has reached its asymptotic velocity, $v_\infty:= c\sqrt{1-1/\Gamma_\infty^2}$, and thus we use this quantity to determine the radial velocity profile instead of the velocity measured at $R_0$.
    From the measured ejecta properties, we determine the time-dependent profiles for the rest-mass density $\rho_{\rm ej}(r,t)$ and velocity $v_{\rm ej}(r,t)$ as described in Appendix \ref{ap:ejecta_modeling}.

    The asymptotic speed of the fastest ejecta layers, $\beta_{\rm max} = \max_{t_0}\left[(v_\infty(t_0)/c \right]$ approximately determines the outermost boundary of the ejecta:
\begin{equation}
    r_{\rm max}(t) = R_0 + \beta_{\rm max}c (t-t_0).
    \label{eq:r_max}
\end{equation}
	However, due to the spread in velocities, the fast ejecta may extend to larger radii and speeds, forming a smooth extended tail for $r>r_{\rm max}(t)$.
	As we will see, the presence or absence of such a tail has a large impact on the properties of the breakout emission.

\begin{figure*}
    \centering
    \includegraphics[width=0.99\linewidth]{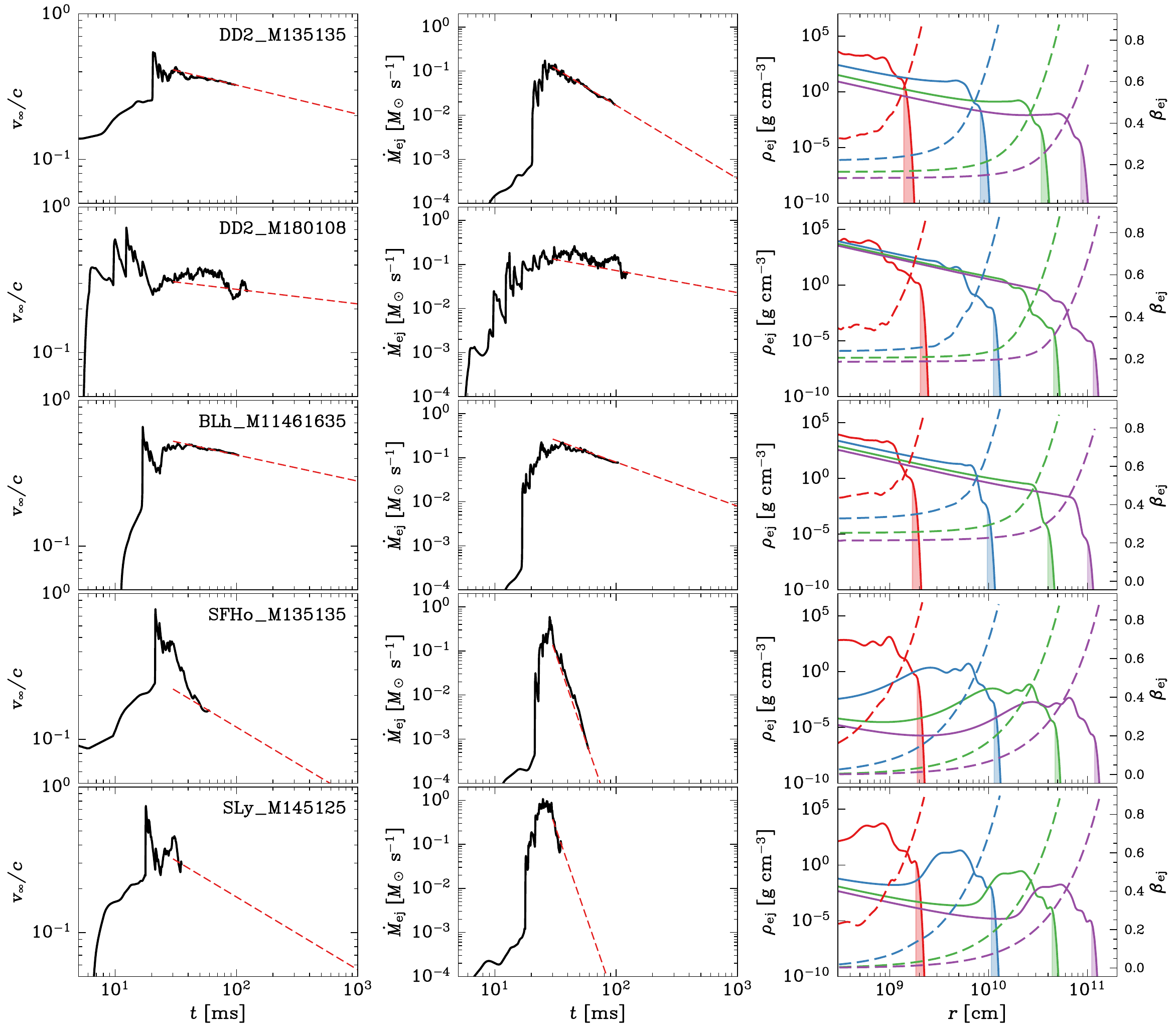}
    \caption{
    Ejecta properties for the five simulations considered (each row).
    The left and middle columns show, respectively, the asymptotic velocity and the mass-loss rate measured at $r=R_0$ as a function of time.
    In red dashed lines, we show the power-law extrapolation of both quantities for times larger than the duration of the simulation.
    The right column shows the outflow rest-mass density (solid lines) and velocity (dashed lines) profiles at four different times: $t=0.1, 0.5, 2, 5~{\rm s}$, denoted by the red, blue, green, and purple curves, respectively.
    The shaded region under the curves corresponds to the extended tail of the ejecta profiles for $r>r_{\rm max}$.
    From top to bottom: DD2\_M135-135, DD2\_M180-108, BLh\_M1146-1635, SFHo\_M135-135, and SLy\_M145-125.}
    \label{fig:profile}
\end{figure*}

	The left and middle panels in Fig. \ref{fig:profile} show the asymptotic velocity and mass-loss rate as a function of time for each simulation.
    The solid dark lines correspond to the quantities measured in the simulation whereas the red dashed lines show the extrapolated power-law profiles.
    The right panel in Fig. \ref{fig:profile} shows the rest-mass density (solid lines) and velocity radial profiles (dashed lines) calculated as described above.
    The different colors correspond to different times after the merger: $t=[0.1,0.5,2,5]~{\rm s}$.
	The extended tail formed by the velocity dispersion in the fast ejecta component corresponds to the shaded regions under each curve.
    
\subsection{Central engine} \label{ssec:engine}

	For a long time, the central engine responsible for launching a sGRB jet has been thought to be a hyperaccreting BH.
	However, recent studies \cite{Mosta_etal2020, CombiSiegel2023b, Kiuchi_etal2024} suggest that the highly magnetized rapidly rotating NS left behind an NS merger may also launch relativistic collimated outflows and thus power sGRBs.
	Depending on the nature of the compact remnant, we assume that one of the two aforementioned central engines launches the relativistic jet.

\subsubsection{Millisecond magnetar} \label{sssec:Magnetar}

	The total rotational energy of a millisecond magnetar with a spin period $P$ is (e.g.,  \cite{Zhang2018pgrb.book.....Z})
\begin{equation}
	E_{\rm rot} \simeq \frac{1}{2} I \Omega^2 \simeq (2.2 \times 10^{52}~{\rm erg}) \left( \frac{M}{1.4 M_\odot} \right) R_6^2 P_{-3}^{-2},
\end{equation}
where
\begin{equation}
	I \simeq \frac{2}{5} M R^2 \simeq ( 1.1 \times 10^{45}~{\rm g~cm}^2 ) \left( \frac{M}{1.4 M_\odot} \right) R_6^2
\end{equation}
is the moment of inertia of the NS (assuming it is spherical), $M$ is its mass, $\Omega = 2\pi / P$ is its angular velocity, $R = (10^6~{\rm cm}) R_6$ is the NS radius, and $P_{-3}=P/(1 ~{\rm ms})$.

	The main mechanism for the energy extraction in millisecond magnetars is magnetic dipole radiation.
	Nevertheless, in the context of NS mergers, other mechanisms such as neutrino- or magnetically-driven winds also play an important role.
	Based on numerical simulations, Ref.~\cite{Siegel_etal2014} derived a semianalytical formula for the Poynting luminosity of a millisecond magnetar formed after the collision of two NSs,
\begin{equation}
	L_{\rm em} \simeq (10^{49}~{\rm erg~s}^{-1} ) B_{{\rm p},15}^2 R_6^3 P_{-3}^{-1},
\end{equation}
where $B=(10^{15}{\rm G}) B_{15}$ is the surface dipolar magnetic field strength at the polar cap region.

	The spin-down timescale $t_{\rm sd}$ can be estimated as
\begin{equation}
	t_{\rm sd} \simeq E_{\rm rot}/L_{\rm em} \simeq (10^3~{\rm s}) \left( \frac{M}{1.4 M_\odot} \right) R_6^{-1} P_{-3}^{-1} B_{{\rm p},15}^{-2}.
\end{equation}
	Since we are interested in timescales $t \ll t_{\rm em}$, it is a fair approximation to consider the magnetar luminosity as constant in time.

\subsubsection{Hyperaccreting black hole} \label{ssseec:BH}

	In early or delayed collapse scenarios, the BH formed may initially have a large spin.
	This BH will be surrounded by a compact, dense, and highly magnetized accretion disk.
	If the magnetic field lines are efficiently advected toward the BH and thread the event horizon, the BH can launch a jet via the Blandford--Znajek (BZ) mechanism \cite{BlandfordZnajek1977, Komissarov2001MNRAS.326L..41K, Komissarov2004MNRAS.350..427K, Tchekhovskoy2011MNRAS.418L..79T}.
    Such a jet will likely be dominated by the Poynting-flux component and its power can be expressed as \cite{Gottlieb_etal2023}
\begin{equation}
    L_{\rm BZ} = \eta_\phi \eta_a \dot{M} c^2,
\end{equation}
where $\eta_\phi \approx (\phi/50)^2$, $\phi:= \Phi_{\rm BH} (R_{\rm g}^2 \dot{M} c)^{-1/2}$ is the dimensionless magnetic flux threading the horizon, $\Phi_{\rm BH}=(1/2)\iint |B^r| \sqrt{-g} d\theta d\phi$ is the dimensional magnetic flux threading the horizon, and $\eta_a \approx 1.063a^4 + 0.395a^2$ parametrizes the jet efficiency associated with the BH spin $a$.
	For a saturated magnetic flux, $\phi \approx 50$, corresponding to a magnetically arrested disk (MAD) state, the BZ power is \cite{Lee_etal2000PhR...325...83L, McKinney2005ApJ...630L...5M, Tchekhovskoy2015ASSL..414...45T}
\begin{equation}
    L_{\rm BZ} \approx 1.7 \times 10^{54} \eta_a \dot{m}~{\rm erg~s}^{-1},
\end{equation}
where $\dot{m} := \dot{M}/(M_\odot{\rm s}^{-1})$.
	This expression gives a power significantly larger than that expected for typical sGRBs.
	
	Numerical simulations have shown that postmerger accreting BHs can launch a jet even before the disk reaches a MAD state \cite{Hayashi_etal2022PhRvD.106b3008H, Hayashi_etal2023PhRvD.107l3001H}.
	This is a direct consequence of the large compactness of the disk, which drives two main effects:
\begin{itemize}
    \item[(a)] The accretion rate starts falling off very rapidly after the merger ($t \gtrsim t_0 \sim 0.1~{\rm s}$), with a power-law time dependence:
\begin{equation}
\label{Mdot}
	\dot{M} \approx \dot{M}_0 (t/t_0)^{-2},
\end{equation}
where $\dot{M}_0 \approx M_{\rm d}/t_0$ and $M_{\rm d}$ is the initial mass of the accretion disk.
    In our case, the two simulations with short-lived remnants (SFHo\_M135-135 and SLy\_M145-125 give rise to accretion disks with a mass $\lesssim 0.01~M_\odot$.
    For simplicity, we fix this parameter to $M_{\rm d}=0.01 M_\odot$ for the scenarios with a BH engine. 
    \item[(b)] The dimensional magnetic flux rapidly accumulates on the BH horizon and finally reaches saturation.
\end{itemize}

	The latter point implies that the jet power remains constant ($L_{\rm BZ} \propto \Phi \sim {\rm const}$) until the time when the disk reaches the MAD state.
	Since the mass accretion rate decays as $\dot{M} \sim t^{-2}$, the magnetic field becomes more relevant with time. This is measured as a growth in the dimensionless magnetic flux until it reaches $\phi \approx 50$.
	After this time, the jet power starts decreasing by following the declining accretion power $L_{\rm BZ} \propto \dot{M} \sim t^{-2}$ [see Eq.~\eqref{Mdot}].

\subsection{Jet and cocoon propagation within the ejecta}

We consider that the central engine launches a magnetized jet with a delay time $t_{\rm del}$ after the merger and a time-dependent luminosity $L_{\rm j}(t)$.
The jet is injected into the merger debris in the polar region at a height $z_0\approx R_0$ with an initial opening angle $\theta_{{\rm j},0}$.
It propagates through the expanding ejecta until it either breaks out or is choked within the ejecta.
As the jet head pushes the ejecta matter, it develops a forward and a reverse shock separated by a contact discontinuity.
    The matter that enters the jet head through the forward shock is heated and pushed sideways, forming a hot cocoon surrounding the jet.
    The jet, initially conical, may be collimated at some height $z_{\rm coll} > z_0$ due to the pressure exerted by the cocoon on its lateral walls.
We assume that the cocoon has an ellipsoidal shape, and as it expands due to its own pressure, it also drives a forward shock onto the ejecta.
   
    The propagation of a jet+cocoon through an ambient medium has been investigated in different setups (e.g., static or expanding medium, magnetized or hydrodynamic jets) by several authors \cite{RamirezRuiz_etal2002, Matzner2003, Morsony_etal2007, Bromberg_etal2011, Salafia_etal2020, HamidaniIoka2021, Gottlieb_etal2023, LevinsonBegelman2013, Bromberg_etal2014,Bhattacharyaetal_2023,Bhattacharyaetal_2024}.
    We here present the system of equations that govern the evolution of the system for our specific problem and leave the full derivation of these equations to Appendix \ref{ap:jet_propagation}.
    We use the subscripts j, h, and c to indicate quantities related to the bulk of the jet, the jet head, and the cocoon, respectively, and the subscript $\perp$ to refer to lateral expansion. 
    The jet head position $z_{\rm h}$, cocoon lateral radius $r_{\rm c,\perp}$, and cocoon internal energy $E_{\rm c}$ evolve as
\begin{eqnarray}
    & \dfrac{dz_{\rm h}}{dt} = \beta_{\rm h} c, \nonumber \\
    & \dfrac{dr_{\rm c,\perp}}{dt} = \beta_{\rm c,\perp} c, \nonumber \\
    & \dfrac{dE_{\rm c}}{dt} = \eta L_{\rm j}(t_{\rm eng}) \left[ \beta_{\rm j} - \beta_{\rm h} \right],
    \label{eq:jet_system}
\end{eqnarray}
where $t_{\rm eng}=t-z_{\rm h}/c$ is the retarded time at the engine, and $\eta$ accounts for the fraction of the jet head that is in causal contact with the cocoon [see Eq. \eqref{eq:eta}] \cite{Salafia_etal2020}.

The breakout time $t_{\rm bo}$ and radius $R_{\rm bo}$ are determined by the condition
\begin{equation}
    \tau = \int_{R_{\rm bo}}^\infty dz \Gamma_{\rm ej}(z) \rho_{\rm ej}(z, t_{\rm bo}) \kappa = 1/\beta_{\rm s}',
    \label{eq:breakout_cond}
\end{equation}
where $\beta_{\rm s}'$ is the forward shock speed in the ejecta frame and $\kappa\simeq 0.16~{\rm cm}^2~{\rm g}^{-1}$ is the absorption coefficient for highly ionized $r$-process elements.

\subsection{Cocoon evolution after the jet head breakout}
\label{sec:cocoon_PBO}

	Once the jet head breaks out at time $t_{\rm bo,h}$ and radius $R_{\rm bo,h}$, it detaches from the ejecta and stops feeding the cocoon.
    However, most of the cocoon material still remains within the ejecta.
    The cocoon evolution is later on driven by the adiabatic conversion of its internal energy to kinetic energy and by its interaction with the remaining ejecta material through which it has to travel to break out.

    Because of its ellipsoidal shape, regions of the cocoon closer to the polar axis will also be closer to the outer boundary of the ejecta and will move faster.
    Then, these regions will likely break out shortly after the jet head breaks out.
    On the contrary, regions that are located at a large polar angle move slower, and thus they may break out with a non-negligible delay with respect to the jet head or even remain trapped within the ejecta.
    Then, in a realistic scenario, it is reasonable to expect a structured cocoon that breaks out with an angular, time, and energy dependence.
	A full treatment of such an evolution requires nonlinear numerical simulations, which are beyond the scope of this paper.
	
	We instead model the cocoon as a single fireball expanding spherically.
	To treat it as a single structure, we first average the cocoon properties over different angles by assuming its energy has an angular distribution given by \cite{Lazzati_etal2017ApJ...848L...6L, LazzatiPerna2019ApJ...881...89L, Salafia_etal2020}:
$dE_{\rm c}/d\Omega \propto  \exp(-\theta/\theta_{\rm c})\cos^2\theta$, where $\theta_{\rm c} \approx r_{\rm c,0}/z_{\rm h,bo}$ is the opening angle of the cocoon.

	To estimate the cocoon expanding speed, we use our knowledge of the speed of the jet head and the cocoon at the time the jet head breaks out as well as its ellipsoidal shape.
    We calculate the horizontal ($x$-directed) and vertical ($z$-directed) cocoon speed components at each polar angle $\theta$ as
\begin{equation}
    \beta_{{\rm c}, x} = \left( \frac{x_{\rm c}}{r_{\rm c,\perp}} \right) \beta_{\rm c,\perp}, ~~~~ \beta_{{\rm c},z}= \left( \frac{z_{\rm c}}{z_{\rm h,bo}} \right) \beta_{\rm h,bo},
\end{equation}
where $x_{\rm c}=r_{\rm c}(\theta) \sin \theta$, $z_{\rm c}=r_{\rm c}(\theta) \cos \theta$, and
\begin{equation}
    r_{\rm c}(\theta) = \frac{z_{\rm h,bo} \cos \theta}{ \cos^2 \theta + \left( \frac{z_{\rm h,bo}}{2r_{\rm c,\perp}} \right)^2 \sin^2\theta }
\end{equation}
gives the radial distance to the origin as a function of $\theta$ for an ellipsoidal surface with semimajor and semiminor axes as $z_{\rm h,bo}/2$ and $r_{\rm c,\perp}$, respectively, and with its center at $(0, z_{\rm h, bo}/2)$.
Finally, the radial expansion speed at a given angle is calculated by (relativistically) adding the projections of $\beta_x$ and $\beta_z$ in the radial direction as
\begin{equation}
    \beta_{\rm c}(\theta) = \frac{\beta_{{\rm c},x} \sin \theta + \beta_{{\rm c},z} \cos \theta}{1+\beta_{{\rm c},x} \beta_{{\rm c},z} \cos \theta \sin \theta}
    \label{eq:beta_ave}
\end{equation}

We then calculate the angle-averaged Lorentz factor and radial coordinate of the cocoon forward shock at the time the jet head breaks out as
\begin{equation}
	\langle \Gamma_{\rm c,0} \rangle \approx \tilde{E}_{\rm c}^{-1} \int_{\Omega_{\rm c}} \left[ 1 - \beta_{\rm c}^2
(\theta) \right]^{-1/2} \frac{dE_{\rm c}}{d\Omega} d\Omega,
 \label{eq:G_ave}
\end{equation}
and
\begin{equation}
    \langle r_{\rm c,0} \rangle \approx \tilde{E}_{\rm c}^{-1} \int_{\Omega_{\rm c}} r_{\rm c}(\theta) \frac{dE_{\rm c}}{d\Omega} d\Omega,
    \label{eq:r_ave}
\end{equation}
respectively, where $\tilde{E}_{\rm c}=\int_{\Omega_{\rm c}} \frac{dE_{\rm c}}{d\Omega} d\Omega$ is the cocoon energy contained within the solid angle $\Omega_{\rm c}$, which extends from the polar axis to the minimum value between $\theta_{\rm c}$ and the angle above which $\beta_{\rm c}(\theta)$ becomes lower than $\beta_{\rm max}$, and thus the matter is likely to remain trapped within the ejecta.

    With the averaged quantities estimated above, we estimate the initial conditions for the cocoon and assume that it propagates as a spherical blast wave.
    The evolution of the Lorentz factor $\Gamma$, the internal energy $U$, the mass $m$, and the radial coordinate $R$ of the blast wave can be obtained by solving the following differential equations \cite{Nava_etal2013}:
\begin{equation}
    \frac{d\Gamma}{dt}= -\frac{\left[ \Gamma - { \Gamma_{\rm ej}} + \Gamma_{\rm eff}{ (\Gamma_{\rm rel}-1)} \right] c^2\frac{dm}{dt} + \Gamma_{\rm eff} \frac{dU_{\rm ad}}{dt}}{(M_{\rm c}+m)c^2 + \frac{d\Gamma_{\rm eff}}{d\Gamma }U},
    \label{eq:dGdt}
\end{equation}
where 
\begin{equation}
    \frac{dm}{dt}=4\pi R^2 \rho_{\rm ej}(R,t) \left( \beta- \beta_{\rm ej} \right) c,
    \label{eq:dmdt}
\end{equation}
is the mass swept up by the forward shock per unit time, $\Gamma_{\rm rel}$ is the Lorentz factor of the cocoon forward shock in the frame of the expanding ejecta material, and $\Gamma_{\rm eff}= (\hat{\gamma}\Gamma^2-\hat{\gamma}-1)/\Gamma$; $\hat{\gamma}\approx 4/3$ is the adiabatic index.
The last term in the numerator is the internal energy loss rate due to adiabatic expansion and can be calculated as
\begin{multline}
    \frac{dU_{\rm ad}}{dt} = -(\gamma_{\rm ad}-1)U\frac{d \ln V'}{dt} = 
    - (\gamma_{\rm ad}-1)U \times \\ \left[ \frac{3 { \beta} c}{R} + \frac{\beta \Gamma \Gamma_{\rm ej}}{\Gamma_{\rm rel}} \frac{d\beta_{\rm ej}}{dt} +
    \left( \frac{\beta_{\rm ej} \Gamma_{\rm ej}}{\Gamma_{\rm rel}\Gamma^2 \beta } - \frac{1}{\Gamma} \right) \frac{d\Gamma }{dt} \right],
\end{multline}
where $V'\sim R^3 \frac{\Gamma_{\rm ej}}{\Gamma_{\rm rel}}$ is the comoving volume of the blast wave.
	In turn, the internal energy of the blast wave evolves as
\begin{equation}
	\frac{dU}{dt} = (\Gamma_{\rm rel}-1)\frac{dm}{dt}c^2 + \frac{dU_{\rm ad}}{dt},
 \label{eq:dUdt}
\end{equation}
where the first term accounts for the shock heating.
The forward shock position evolves as
\begin{equation}
	\frac{dR}{dt}=\beta c,
 \label{eq:drdt}
\end{equation}
where $\beta$ is obtained from the Lorentz factor.

	We solve the system of Eqs. \eqref{eq:dGdt}, \eqref{eq:dmdt}, \eqref{eq:dUdt}, and \eqref{eq:drdt} with initial conditions $(\langle \Gamma_{\rm c,0} \rangle, M_{\rm c}, \tilde{E}_{\rm c}, \langle r_{\rm c,0} \rangle)$ to determine the cocoon breakout properties [the cocoon breakout is also calculated using Eq. \eqref{eq:breakout_cond}].
 Here, $M_{\rm c}$ is the ejecta mass enclosed within the cocoon volume at the time when the jet head breaks out.

\section{Results on jet and cocoon propagation} \label{sec:results}

    To explore the parameter space of relevance in our models, we have performed $200$ simulations of the jet+cocoon evolution for each ejecta profile and each of the two approaches to model its outermost layers (extended tail or sharp cutoff) by varying the following central engine parameters: the dipolar magnetic field $B$ threading the engine (either the event horizon if the engine is a BH or the NS surface if the engine is a millisecond magnetar), the time delay between the merger and the jet launch $t_{\rm del}$, and the initial opening angle of the jet $\theta_{\rm j,0}$.
    We choose the first two parameters to be uniformly distributed in the ranges $B \in (5 \times 10^{14}, 10^{16})~{\rm G}$ and $t_{\rm del} \in (0.1, 2)~{\rm s}$, respectively, whereas for the jet opening angle, we consider two different values, $\theta_{\rm j,0} = 6^\circ,~18^\circ$, as representative cases of ``narrow'' and ``wide'' jets, respectively.
	In what follows, we analyze and compare for each model the jet and cocoon properties at the time of breakout of each of both components.

\begin{figure*}
    \centering
    \includegraphics[width=0.98\linewidth]{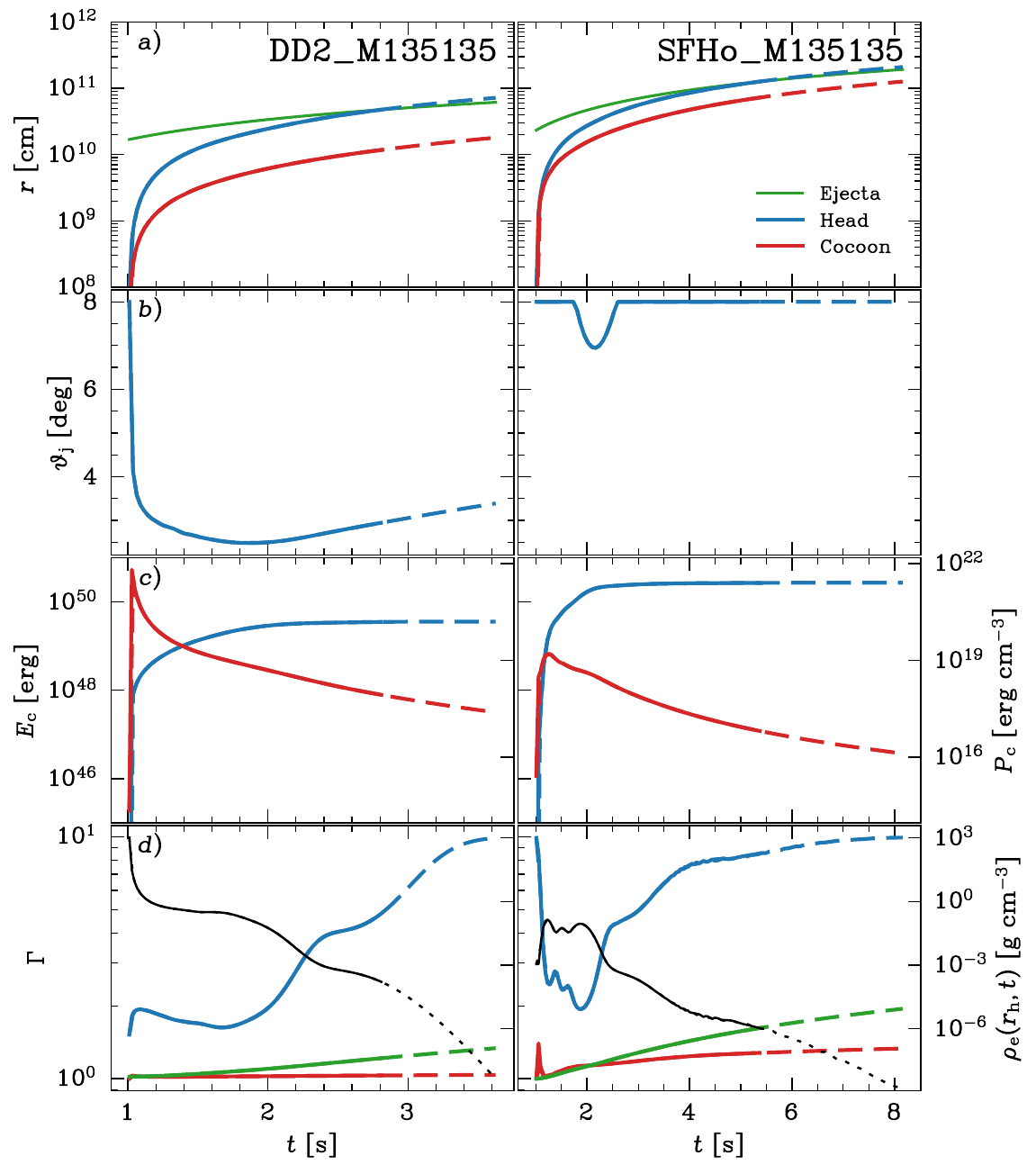}
    \caption{Time evolution of the jet head and the cocoon for two different simulations and engine models. Left: DD2\_M135-135 simulation with a magnetar engine. Right: SFHo\_M135-135 simulation with a BH engine.
    Both engine models have an initial jet opening angle of $\theta_0=8^\circ$, a dipolar magnetic field $B=5\times 10^{15}~{\rm G}$, and a jet launching time delay of $t_{\rm del}=1~{\rm s}$.
    From top to bottom: {\it a)} position of the jet head, the outer boundary of the ejecta and lateral radius of the cocoon; {\it b)} opening angle of the jet; {\it c)} cocoon energy (blue) and cocoon pressure (red); and {\it d)} Lorentz factor for the jet head (blue), the outer boundary of the ejecta (green), and the lateral radius of the cocoon (red). The black line shows the rest-mass density at the jet head position as a function of time.
    For all the panels, the dashed extended lines correspond to the evolution of the quantities when the ejecta density profile has an extended tail.}
    \label{fig:jet_propagation_1}
\end{figure*}
    
	In Fig.~\ref{fig:jet_propagation_1}, we show the time evolution of the jet head position (a), the jet opening angle (b), the internal energy and pressure of the cocoon (c), and the Lorentz factor of the jet head (d) for the ejecta profiles derived from DD2\_M135-135 with a magnetar engine (left column panels) and SFHo\_M135-135 with a BH engine (right column panels). 
 Figure \ref{fig:jet_propagation_1}(a) also shows the radius of the ejecta outer boundary [Eq. \eqref{eq:r_max}] and the lateral radius of the cocoon, while Fig. \ref{fig:jet_propagation_1}(d) shows their respective Lorentz factors.
	In both examples, the engine has a dipolar magnetic field $B=5\times 10^{15}~{\rm G}$ and launches a jet with an initial opening angle of $\theta_{\rm j,0}=8^\circ$ and with a time delay of $t_{\rm del}=1~{\rm s}$ after the merger.\footnote{Note, however, that the same parameters give a larger jet power for a BH engine than for a magnetar engine.}

	The ejecta from SFHo\_M135-135 is less massive and expands faster than the one derived from DD2\_M135-135.
	This causes the cocoon to expand laterally much faster, as can be noticed by comparing the two columns in panel d) of Fig.~\ref{fig:jet_propagation_1}.
	The cocoon then acquires a larger volume, which implies that it exerts a lower pressure on the lateral walls of the jet compared to the case of DD2\_M135-135, despite having a larger internal energy.
	Consequently, the jet is collimated only for a short time, after which the jet expands back to attain the same opening angle with which it was initially launched ($8^\circ$), as shown in the right column of panel b) for $\theta_j$.

	The solid lines in all four plots correspond to the cases where the ejecta profile has a sharp cutoff at $r=r_{\rm max}(t)$, while the dashed lines show the continuation of the jet+cocoon evolution when the ejecta has a smooth extended tail at $r>r_{\rm max}(t)$.
	For the ejecta profile with the sharp cutoff, the breakout condition given by Eq. \eqref{eq:breakout_cond} is almost interchangeable with $R_{\rm bo} \approx r_{\rm max}(t_{\rm bo})$, whereas for the profiles with an extended tail, we generally have $R_{\rm bo} > r_{\rm max}(t_{\rm bo})$.
	This can be seen in Fig.~\ref{fig:jet_propagation_1}(a), where the dashed lines extend beyond the theoretical outer boundary of the ejecta defined by Eq.~\eqref{eq:r_max}; this fact naturally delays the breakout time.
	
 Another key difference between the two ejecta profiles considered can be observed in the maximum Lorentz factor achieved by the jet head at break out.
	With the sharp cutoff in the ejecta density profile, the jet for SFHo\_M135-135 breaks out with a larger Lorentz factor than in the case of DD2\_M135-135 due to the differences in the density and expansion speed of the ejecta outer layers between the two profiles.
	However, when a fast extended tail is included, both jet heads keep accelerating and break out with a similar Lorentz factor.
	Although the total mass in the extended tail is much lower than the total mass of the ejecta, $m_{\rm ej}(r>r_{\rm max}(t)) \ll M_{\rm ej}(t)$, this component may have a significant effect on the late-time evolution of the jet head by delaying the breakout time and allowing the jet head to reach larger speeds.

\begin{figure}
    \centering
\includegraphics[width=0.95\linewidth]{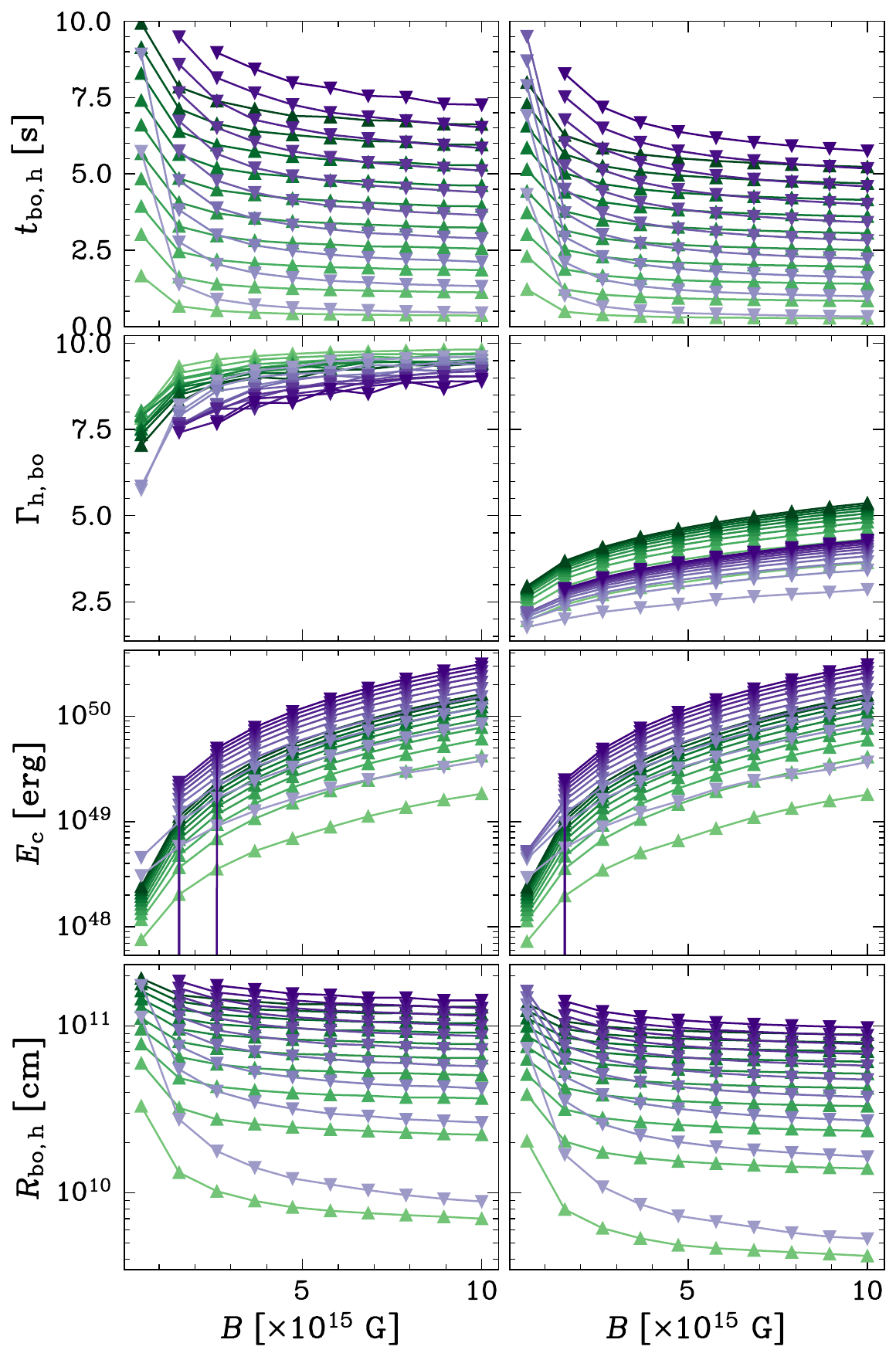}
	\caption{Properties of the jet+cocoon system at the time of the jet head breakout as a function of the dipolar magnetic field at the engine for the simulation DD2\_M135-135. Left: ejecta profile with an extended tail. Right: ejecta profile with a sharp cutoff at $r=r_{\rm max}(t)$. From top to bottom: jet breakout time in the lab frame ($t_{\rm bo,h}$), Lorentz factor of the jet head ($\Gamma_{\rm h,bo}$), cocoon internal energy ($E_{\rm c}$), and jet head breakout radius ($R_{\rm bo,h}$). The green (purple) lines show the results from simulations where the engine launches a jet with an initial opening angle of $\theta_{\rm j,0}=8^\circ$ ($\theta_{\rm j,0}=16^\circ$). For both these cases, each individual curve corresponds to a fixed time delay $t_{\rm del} \in (0.1, 2)~{\rm s}$ with a darker line color corresponding to a larger delay in jet launching.}
	\label{fig:jet_DD2_M135135}
\end{figure}
    
	From the $200$ engine models considered for each ejecta profile, we derive the main properties of the jet head and the cocoon at the breakout time: the jet head location $z_{\rm h, bo} \equiv R_{\rm bo,h}$, the jet head speed $\beta_{\rm h,bo}$, the lateral radius of the cocoon $r_{\rm c,\perp}$ and its lateral expansion speed $\beta_{\rm c,\perp}$, as well as the cocoon internal energy $E_{\rm c}$.
	As an example, in Fig. \ref{fig:jet_DD2_M135135} we show the properties of the jet+cocoon system at the breakout time for the ejecta profile derived from the simulation DD2\_M135-135 for the range of engine parameters considered.
	The left (right) panels correspond to the scenario with a sharp cutoff (extended tail) in the ejecta profile.
	Each curve shows the respective quantity as a function of the dipolar magnetic field $B$ of the engine, for fixed values of $(\theta_{\rm j,0}, t_{\rm del})$.
	We discriminate between the two initial opening angles of the jet with different colors, using green (purple) for $\theta_{\rm j,0}=8^\circ$ ($\theta_{\rm j,0}=16^\circ$).
	For fixed values of $B$ and $t_{\rm del}$, jets launched with a smaller opening angle have a larger Lorentz factor at breakout, a shorter breakout time and radius, and they deposit a smaller amount of energy into the cocoon.
	In addition, the presence of an extended tail in the ejecta profile (see left panels) induces larger Lorentz factors at the time of breakout, since the jet head has more time to accelerate while traveling through the ejecta outer layers.

\begin{figure}
    \centering
\includegraphics[width=0.99\linewidth]{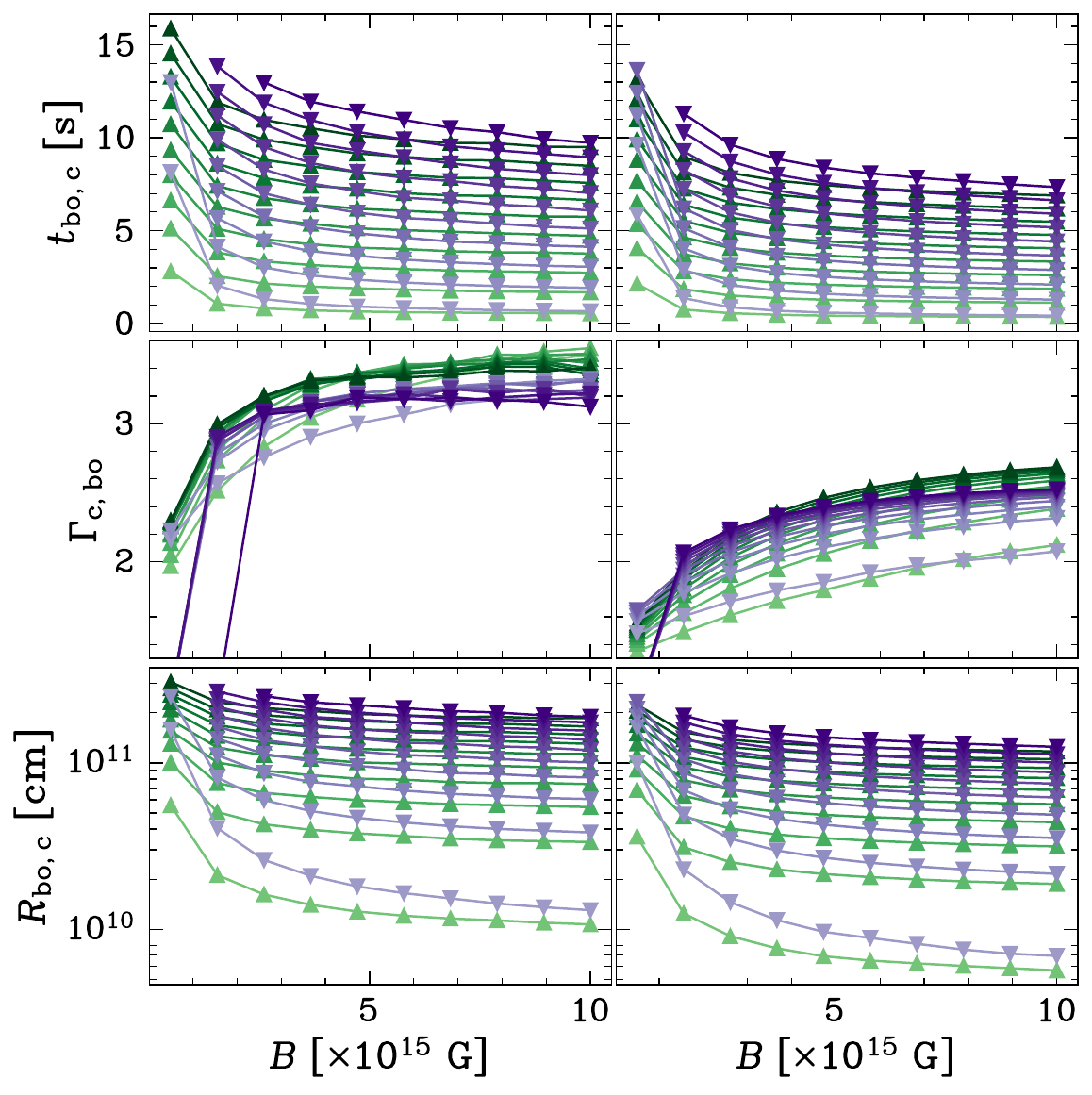}
	\caption{Similar to Fig. \ref{fig:jet_DD2_M135135} but for the cocoon breakout. From top to bottom: cocoon breakout time ($t_{\rm bo,c}$), Lorentz factor of the cocoon at breakout ($\Gamma_{\rm bo,c}$), and cocoon breakout radius ($R_{\rm bo,c}$).}
	\label{fig:cocoon_DD2_M135135}
\end{figure}

\begin{figure}
    \centering
\includegraphics[width=0.999\linewidth]{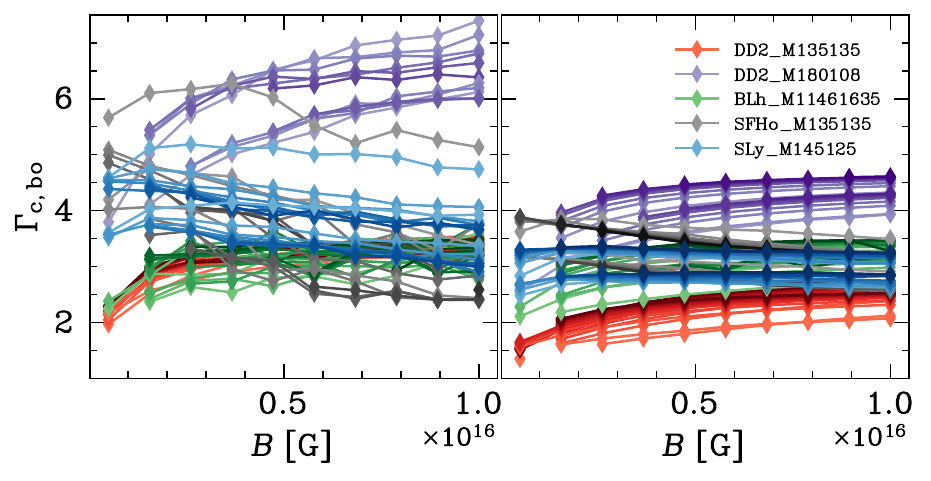}
    \caption{Cocoon Lorentz factor at the time of its breakout as a function of the dipolar magnetic field powering the jet. Here we show the results for all five NR simulations considered, each denoted by a different color. Each curve corresponds to a fixed value for the jet opening angle and the jet launching time delay, with a darker color corresponding to a larger delay. Left: Ejecta profile with an extended tail. Right: Ejecta profile with a sharp cutoff.}
    \label{fig:cocoon_Gamma}
\end{figure}

	For each simulation, the values of the quantities we derived at the time of jet head breakout serve as the initial conditions for the subsequent cocoon evolution, which is determined as described in Sec.~\ref{sec:cocoon_PBO}.
	Fig.~\ref{fig:cocoon_DD2_M135135} shows the properties of the cocoon at the time it breaks out.
	The dependence of these properties on the engine parameters is similar to that of the jet head quantities shown in Fig. \ref{fig:jet_DD2_M135135}.
	The main difference with the jet head is the smaller overall Lorentz factor that the cocoon reaches at breakout; this is the case for profiles with a sharp cutoff and those with an extended tail.
	This is reasonable given the larger opening angle of the cocoon and the lack of continuous energy injection from the engine once the jet head has detached from the cocoon.
	Fig.~\ref{fig:cocoon_Gamma} shows the maximum Lorenz factor achieved by the cocoon at the time of its breakout for all five simulations considered:
	DD2\_M180-108 shows the largest values and DD2\_M135-135 has the lowest values for $\Gamma_{\rm c,bo}$.
 
	Simulations producing an early collapse and thus a BH engine show a slightly negative correlation between the Lorentz factor and the engine magnetic field, as opposed to the simulations with a long-lived NS engine.
	This is a consequence of BH engines being more powerful (by about an order of magnitude) than magnetar engines for the same magnetic field strength.
    For large (small) jet powers, an increase in the dipolar magnetic field tends to cause an increase (a decrease) in the lateral radius of the cocoon at the time when the jet head breaks out. 
    In turn, a larger lateral radius of the cocoon implies a larger cocoon opening angle and lower average cocoon Lorentz factor.
    Then, we see opposite trends for BH engines (large jet powers) and magnetar engines (small jet powers).

\section{Shock breakout emission}
\label{sec:shockbreakout}

	The theory of Newtonian shock breakout and its associated electromagnetic signal have been extensively studied by many authors in the context of static stellar envelopes, such as those occurring in core-collapse supernovae \cite{KleinChevalier1978ApJ...223L.109K, EnsmanBurrows1992ApJ...393..742E, Sapir_etal2011ApJ...742...36S, Katz_etal2010ApJ...716..781K, Katz_etal2012ApJ...747..147K, NakarSari2010ApJ...725..904N, Piro_etal2010ApJ...708..598P} or in stellar winds \cite{Ioka_etal2019MNRAS.484.3502I, Ito_etal2020MNRAS.499.4961I}.
	Relativistic shock breakout theory was first explored in Ref. \cite{NakarSari2012} for static envelopes and later investigated for expanding media \cite{Kasliwal_etal2017, Gottlieb_etal2018a, Gottlieb_etal2018b, Bromberg_etal2018MNRAS.475.2971B, Nakar2020}; see also Ref. \cite{Ito_etal2020MNRAS.492.1902I}.
	The emission produced in a shock breakout from expanding ambient media may largely depend on the properties of the ejecta profile at the outermost boundaries and present significant differences from that produced in a static medium \cite{Nakar2020}.
    In what follows, we summarize the main steps involved in calculating such emission.

\subsection{General estimates}

	To simplify the notation, we use primed notation to refer to quantities that are measured in the ejecta frame\footnote{Note that there is not a unique ejecta frame (it expands with a distribution of speeds); hence, when we use primed quantities, we are referring to the ejecta frame at a specific radius and time.}: e.g., $\Gamma_{X{\rm ej}} \equiv \Gamma_X'$.
	The forward shock driven by the jet or the cocoon propagating through the ejecta in NS mergers will likely be radiation mediated (dissipation within the transition layer is mediated by radiation) and radiation dominated (the downstream internal energy is dominated by radiation, implying an adiabatic index $\gamma_{\rm ad}\approx 4/3$).
	The shock propagates as long as its optical depth $\tau$ remains greater than $\sim 1/\beta_{\rm s}'$ [see Eq. \eqref{eq:breakout_cond}]; after this time, the photons within the shock transition layer can diffuse out, and we can say that the shock breaks out.
	These first escaping photons arise from the so-called ``breakout layer'', and are then followed by photons generated deeper in the downstream, which need a longer diffusion time to escape the surrounding medium.

	The condition for a shock to break out can be expressed as $\tau \sim \kappa m_{\rm bo,min}/(4\pi R_{\rm bo}^2) \approx 1/ \beta_{\rm s}'$, where $m_{\rm bo, min}$ is the minimum mass in the breakout layer,
\begin{equation}
    m_{\rm bo, min} \approx \frac{4\pi R_{\rm bo}^2}{\beta_{\rm s}' \kappa}.
\end{equation}
    The downstream is energized due to the deceleration of the material crossing the shock, and this energy is radiated after the breakout.
    The internal energy of the breakout layer in the lab frame can be estimated as
\begin{equation}
    E_{\rm bo,min} \sim \gamma_{\rm bo} (\gamma_{\rm bo}' - 1) m_{\rm bo,min} c^2,
\end{equation}
where $\gamma_{\rm bo}$ is the Lorentz factor of the breakout layer.
The right panel of Fig. \ref{fig:cartoon} shows schematically the basic picture described above.

	The duration of the breakout signal can be approximated by the ``angular time scale'', defined as the difference between the light travel time of photons emitted along the line of sight and that of photons emitted at an angle $\sim 1/\gamma_{\rm bo}$. Thus, $\Delta t_{\rm bo} \sim R_{\rm bo} / (c \gamma_{\rm bo}^2)$.
	After the first photons from the breakout layer escape, radiation starts to diffuse out of deeper layers.
	The phase comprised between the time since the release of the first photons, i.e. the breakout time, until the time when the expanding gas doubles its radius is usually known as the ``planar phase''.
	If the shocked gas is relativistic ($\gamma_{\rm s} \beta_{\rm s} \gtrsim 1$), the dynamical timescale $\sim R_{\rm bo}/(\beta_{\rm s} c\gamma_{\rm s}^2)$ which determines the duration of the planar phase is comparable to the angular timescale. Consequently, photons that diffuse out during this phase can arrive simultaneously to the observer with those emitted by the breakout layer.
 The net result of this effect is an energy enhancement of the received signal \cite{Gottlieb_etal2018b, Nakar2020}.

    The degree to which the signal is enhanced depends largely on the shape of the ejecta profile close to its outermost edge.
	To carefully quantify this effect, we seek for the radius $R_{\rm em}$ from which the photons diffuse out over the duration of the shock breakout signal \cite{Gottlieb_etal2018a, Gottlieb_etal2018b}.
    To to this, we equalize the diffusion timescale between $R_{\rm em}$ and the photospheric radius $R_{\rm ph}$ with the dynamical timescale in the comoving frame $t'_{\rm dyn} \sim R_{\rm bo} / (\gamma_{\rm bo}c)$ \cite{GinzburgBalberg2012ApJ...757..178G},
 \begin{eqnarray}
     t'_{\rm diff}(R_{\rm em}) = \frac{1}{c}\int_{R_{\rm em}}^{R_{\rm ph}} d(r-R_{\rm em})^2 \gamma^2 \kappa \rho_{\rm ej}'(r, t) \nonumber \\
     \approx \frac{2 \kappa}{c}\int_{R_{\rm em}}^{R_{\rm ph}} dr (r-R_{\rm em})\gamma^2 \rho_{\rm ej}'(r, t).
     \label{eq:t_diff}
 \end{eqnarray}
We then estimate the total mass contributing to the radiation during this period as $m_{\rm bo}\approx 4\pi \int_{R_{\rm em}}^{R_{\rm ph}} dr\ r^2\rho_{\rm ej}(r, t)$, while the energy released is \cite{Nakar2020}
\begin{equation}
	E_{\rm bo} \simeq E_{\rm bo,min} \frac{m_{\rm bo}}{m_{\rm bo,min}}.
 \label{eq:E_bo}
\end{equation}

	The total energy radiated thus depends largely on the specific profile of the ejecta outermost layers.
	If the profile has a sharp cutoff, then the density is large near the edge and $R_{\rm ph}(t) \approx r_{\rm max}(t)$.
	As a result, a large amount of mass may contribute during the planar phase and $m_{\rm bo} \gg m_{\rm bo,min}$.
	In contrast, if the profile has an extended low-density tail, the breakout may occur before reaching the outermost layers in a region with a lower density such that $m_{\rm bo} \gtrsim m_{\rm bo, min}$.
    To be more precise, let us assume that the radiating layer has a width $\Delta'$ and an associated diffusion time $t'_{\rm diff} \propto \Delta'^2 \rho'$; the radiating mass contained within that layer is $m\propto \Delta'\rho'$.
    Profiles with an extended tail have a lower mass density at the breakout radius than those with a sharp cutoff, namely $\rho'_{\rm ET} \ll \rho'_{\rm SC}$ (we used the subscripts ET and SC for quantities in the extended tail and the sharp cutoff scenarios, respectively).
    Then, for a fixed diffusion time, the ratio of widths for the two types of profile is $\Delta'_{\rm ET}/\Delta'_{\rm SC} \propto \sqrt{\rho'_{\rm SC} / \rho'_{\rm ET}}$, and the layer masses are related as $m_{\rm ET} \propto \rho'_{\rm ET} \Delta'_{\rm ET} \sim \sqrt{\rho'_{\rm ET} \rho'_{\rm SC}} \Delta'_{\rm SC} \sim \sqrt{\rho'_{\rm ET}/\rho'_{\rm SC}} m_{\rm SC}$.
        Since the radiated energy is proportional to the mass in the radiating layer, we have $E_{\rm bo, ET} \ll E_{\rm bo, SC}$.

	From the simulations described in the previous section and the above analysis, we can straightforwardly derive two observables: the isotropic equivalent energy radiated during the breakout $E_{\rm bo}$ and the time delay between the arrival of GWs and that of the $\gamma$ rays, which in our case is given by the observed breakout time
\begin{equation}
    t_{\rm bo, obs} = t_{\rm bo,c} - \frac{R_{\rm bo,c}}{c}.
    \label{eq:t_bo_obs}
\end{equation}
 In addition, we can extract information about the average energy of photons in the observer frame as follows.
	It is known that if the downstream temperature exceeds ${\sim}50 ~{\rm keV}$, electron-positron pair production becomes important and significantly affects the shock structure.
	In particular, the temperature in the immediate downstream is regulated by pair creation and annihilation, and is maintained in the range ${\sim} 100{-}200~{\rm keV}$ \cite{Katz_etal2010ApJ...716..781K, Budnik_etal2010ApJ...725...63B}.
	A large number of pairs can significantly increase the effective optical depth, thereby delaying the photon escape until the gas expands and the temperature drops back to ${\sim} 50~{\rm keV}$.
	This implies that, independent of the specific shock details, it can be expected that the breakout photons are released with a temperature of ${\sim} 50~{\rm keV}$ in the downstream frame \cite{NakarSari2012, Granot_etal2018MNRAS.476.5453G, Lundman_etal2018ApJ...858....7L}.
	The observed temperature can then be approximated by \cite{NakarSari2012}
\begin{equation}
    T_{\rm obs} \sim 50 \gamma_{\rm bo,f} ~{\rm keV},
\end{equation}
where $\gamma_{\rm bo,f}$ is the Lorentz factor of the emitting plasma after it has cooled down to $50~{\rm keV}$.
	In principle, this value can be larger than the Lorentz factor the plasma acquired through the shock passage.
	If the shock is relativistic with $\gamma'_{\rm s} \beta_{\rm s}' \gtrsim 1$, this acceleration occurs before the photons are released, since the pair opacity will be very high.
	The final Lorentz factor in the ejecta frame is $\gamma_{\rm bo,f}' \approx \gamma_{\rm bo}'^\alpha$, where $\alpha$ is in the range $(1, 1+\sqrt{3})$ \cite{JohnsonMcKee1971PhRvD...3..858J, PanSari2006ApJ...643..416P}.
 Because of the uncertainties in this parameter, we neglect the postbreakout acceleration and assume $\alpha\approx 1$.

\subsection{Bolometric light curve}
\label{subsec:lightcurve}

	In this subsection, we calculate the bolometric light curve for the shock breakout emission derived from our model and the spectrum assuming purely thermal emission.
    To this aim, we determine the equal-arrival-time surface (EATS) \cite{WoodsLoeb1999ApJ...523..187W, Salafia_etal2016MNRAS.461.3607S} for each observer time $t_{\rm obs} > t_{\rm bo, obs}$ and integrate the emission over it.
    The EATS is determined by the equation
\begin{equation}
	t_{\rm obs} = t - \frac{R(t)}{c} \cos \theta,
\end{equation}
where $\theta$ is the polar angle measured from the line of sight, and $t$ is the time in the lab frame.
	At time $t$, the emission comes out from the photospheric radius $R_{\rm ph}(t)$, which is determined by solving $\int_{R_{\rm ph}(t)}^\infty \Gamma(r,t) \rho_{\rm ej}(r,t) \kappa dr = 1$.
 	However, this radiation is originally produced in a deeper layer at $R_{\rm em}(t)$ such that the time it takes for the radiation to diffuse up to the photosphere equals the time elapsed after the shock breakout, measured in the comoving frame.
  This is, $t'_{\rm diff}(R_{\rm em}, R_{\rm ph}) \approx (t-t_{\rm bo})/ \langle \Gamma_{\rm ej} \rangle$, where
$t_{\rm diff}'$ is given by Eq. \eqref{eq:t_diff}, $\langle \Gamma_{\rm ej} \rangle$ is the average Lorentz factor of the ejecta between $R_{\rm em}(t)$ and $R_{\rm ph}(t)$, and we have assumed that diffusion occurs in the radial direction.
	Once we obtain the lab frame time $t$ and the emission radius $R_{\rm em}(t)$ for each $\theta$, we calculate the isotropic equivalent luminosity at the observer time $t_{\rm obs}$ as $L_{\rm iso}=4\pi d_L^2 F$, where $d_L$ is the luminosity distance to the source and $F$ is the bolometric flux:
\begin{equation}
	F(t_{\rm obs}) = \frac{1}{4\pi d_L^2} \int_{\mu_{\rm on}(t_{\rm obs})}^1 \mathcal{D}^4 \frac{dL'(t)}{d\mu} \mu d\mu.
\end{equation}
Here, $\mu = \cos \theta$, $\mathcal{D}= [ \Gamma (1-\beta \mu ) ]^{-1}$ is the Doppler factor of the fluid moving at an angle $\theta$ with respect to the observer, and $dL'/d\mu$ is the comoving luminosity per unit $\cos \theta$ at time $t$.
Since the emission escapes only for $t>t_{\rm bo,c}$, at any time the observer sees the emission coming from regions up to the maximum angle $\theta_{\rm on}$ beyond which $t(\theta_{\rm on})<t_{\rm bo,c}$; this gives $\mu_{\rm on}(t_{\rm obs})=1-c(t_{\rm obs}-t_{\rm bo, obs})/R_{\rm bo,c}$ (see, e.g., Ref. \cite{Salafia_etal2016MNRAS.461.3607S}).

To calculate the energy released by the shocked plasma per unit time, we consider the following argument.
At a given time $t_1>t_{\rm bo}$, the emission escaping from the photosphere has to be produced in a deeper layer located at the correspondent emission radius: $R_1 = R_{\rm em}(t_1)$.
Since this layer expands at a speed $\beta_1 = \beta_{\rm ej}(R_1, t_1)$, at a later time $t_2 = t_1 + \Delta t$, this layer will have expanded to a radius $R_2 = R_1 + \beta_1 c \Delta t$, while the new emission radius is $R_{\rm em}(t_2) < R_2$.
That means that during the interval $\Delta t$, the internal energy contained between $R_{\rm em}(t_2)$ and $R_2=R_{\rm em}(t_1) + \beta_1 c \Delta t$ has been radiated away.
If we sit in the comoving frame of a given layer, its radius remains constant while the emission radius $R_{\rm em}'$ drifts inward.

Hence, we estimate the comoving luminosity as
\begin{equation}
    \frac{dL'}{d\mu}(t) \approx 2\pi R_{\rm em}^2(t) \frac{4}{3}e'(t,R_{\rm em}) \left| \dot{R}'_{\rm em}(t) \right|,
\end{equation}
where $e'(t, R_{\rm em})$ is the postshock thermal energy density in the ejecta, which is obtained by assuming that the shocked ejecta layers expand adiabatically after the shock passage and until they radiate away their energy, and $| \dot{R}'_{\rm em}(t) |$ gives the rate at which the emission radius drift inward in the ejecta comoving frame.
This is given by the relativistic transformation of velocities: $\dot{R}'_{\rm em}(t) = [\dot{R}_{\rm em}(t) - \beta_{\rm ej}c] / [ 1 - \beta_{\rm ej}\dot{R}_{\rm em}(t)/c]$, where $\beta_{\rm ej} = \beta(R_{\rm em}, t)$.

\begin{figure}
	\includegraphics[width=0.94\linewidth]{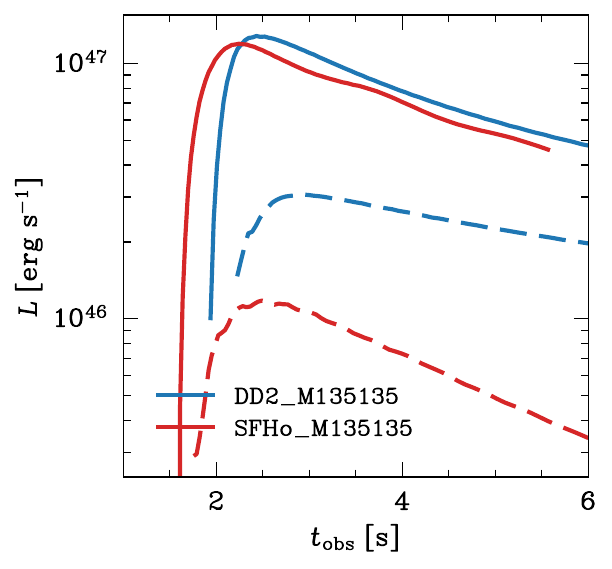}
 \caption{Bolometric light curve for radiation emitted at the cocoon shock breakout for the four evolution scenarios shown in Fig. \ref{fig:jet_propagation_1}.
 These correspond to engines (magnetar for DD2\_M135-135 and BH for SFHo\_M135-135) with parameters $B=5\times 10^{15}~{\rm G}$, $t_{\rm del}=1~{\rm s}$, and $\theta_{\rm j,0}=8^\circ$.
 Solid (dashed) lines correspond to the case where the ejecta profile has a sharp cutoff (extended tail) at its outermost boundary.}
 \label{fig:lightcurve}
\end{figure}

Figure \ref{fig:lightcurve} shows the bolometric light curves (top panel) for the two reference cases analyzed in Fig. \ref{fig:jet_propagation_1}.
The blue curves show the emission from the cocoon breakout for the ejecta model where the ejecta is derived from DD2\_M135-135 and the engine is a magnetar, whereas the red curves correspond to the case where the ejecta is derived from SFHo\_M135-135 and the engine is an accreting BH.
For both scenarios, the engine has a dipolar magnetic field $B=5\times 10^{15}~{\rm G}$, and the jet is launched with a time delay $t_{\rm del}=1~{\rm s}$ and an opening angle $\theta_{\rm j,0}=8^\circ$.
The shock breakout emission of ejecta profiles with a sharp cutoff (solid lines) shows larger luminosities, up to an order of magnitude above those corresponding to the extended tail scenarios.

\subsection{Spectrum}

   The spectral shape of relativistic shock breakout emission is largely unknown.
    Although the spectrum can have a thermal component, at high energies it may resemble an approximate power law \cite{NakarSari2012}; see also Refs. \cite{Ito_etal2020MNRAS.492.1902I, Ito_etal2020MNRAS.499.4961I}.
    Particle acceleration is not expected to be efficient in radiation-mediated shocks, but a power-law energy component can still arise by other mechanisms, such as, e.g., Comptonization of the radiation by the less dense outermost layers, light-travel-time effects which lead to photons from different radii and different angles arriving together at the observer, subshocks, or more complex effects such as the so-called photon acceleration \cite{GruzinovMeszaros2000ApJ...539L..21G}.
    In addition, the temperature of the deeper ejecta layers that radiate after the breakout layer may have a rest-frame temperature lower than $50$~{\rm keV}, which could shift the peak to lower energies and modify the spectral shape above the peak \cite{Gottlieb_etal2018b, Nakar2020}.

    In this work, we do not aim to explore in detail the spectral shape of the emitted radiation.
    We conservatively show what the spectrum looks like if we assume that the energy radiated has a pure Wien (thermal) spectrum and the layers involved in the first few seconds after shock breakout have a temperature of $50~{\rm keV}$ in their rest frame.
	The total observed spectrum at a given time is then the superposition of several Wien spectra with different effective temperatures $T_{\rm eff}(\theta) \sim \mathcal{D}(\theta) {\times} 50~{\rm keV}$ corresponding to rings in the range $(\theta, \theta{+}d\theta)$.
    The luminosity emitted from each ring is $\propto \frac{\sin \theta}{(1-\beta \cos \theta)^3}$.
    
\begin{figure}
 \includegraphics[width=0.94\linewidth]{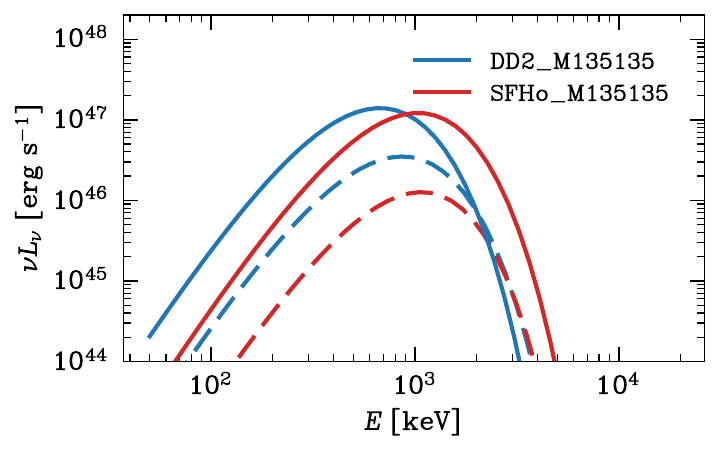}
 \caption{Thermal spectrum at $t=1.1 t_{\rm bo, obs}$ for the same scenarios described in Fig. \ref{fig:lightcurve}.}
 \label{fig:spectrum}
\end{figure}

Figure \ref{fig:spectrum} shows spectra for the same four scenarios described in Sec. \ref{subsec:lightcurve}
The peak energies are at $E_{\rm pk} \sim 2.7 \Gamma_{\rm bo} \times (50 ~{\rm keV})$ due to the assumption of a purely thermal spectrum.
The resultant spectra are similar to a single Wien spectrum, though with a small excess at energies below the peak due to the contribution of the larger angle of view regions that are less beamed toward the observer.
As we mentioned above, if the temperature in the deeper layers is lower than $50~{\rm keV}$ in its rest frame, the peak could be shifted to lower energies, and the spectrum would differ even more from a Wien spectrum.

\section{Application to GW170817}
\label{sec:GW170817}
GW170817 is the only event to date that has been detected with multi-messenger emission comprising both GWs and EM radiation.
    The GRB signal arrived with a time delay of $t_{\rm GRB}=(1.73 \pm 0.05)~{\rm s}$ with respect to the GWs, and the main pulse had a peak energy of $E_{\rm pk,GRB}=(185 \pm 65)$ keV and an isotropic equivalent energy of $E_{\rm GRB}=(5.1 \pm 1)\times 10^{46}$ erg.
	In our model, the detected emission results from the cocoon shock breakout, and so the observed delay between the GWs and the GRB corresponds to the observed cocoon breakout time [Eq. \eqref{eq:t_bo_obs}].
	This includes both the potential delay at the engine for the launching of the jet and the propagation time for the jet and the cocoon within the ejecta.
 Similarly, if the comoving temperature of the emitting plasma is ${\sim} 50~{\rm keV}$, then the peak energy should be approximately given by $E_{\rm pk,GRB}\sim \gamma_{\rm bo} \times 50$ keV, and thus the observed peak energy provides information about the Lorentz factor of the emitting plasma.

 In what follows, we apply two different approaches to compare the predictions of our model with the data from GRB 170817A.

\subsection{Null test for the ejecta profiles}
\label{subsec:bayesian_I}

	Assuming that the breakout radius is approximately given by the outermost ejecta boundary, $R_{\rm bo,c} \approx R_0 + \beta_{\rm e, max}c t_{\rm bo,c} \approx \beta_{\rm e, max}c t_{\rm bo,c}$, the cocoon breakout occurs in the engine (lab) frame at a time [see Eq. \eqref{eq:t_bo_obs}]
\begin{equation}
    t_{\rm bo,c} \approx \frac{t_{\rm bo, obs}}{1-\beta_{\rm e, max}} \approx \frac{t_{\rm GRB}}{1-\beta_{\rm e, max}}.
\end{equation}
	Then, for a given ejecta model (we have $\beta_{\rm e,max}$), we use the known values for $t_{\rm GRB}$ and $E_{\rm pk,GRB}$ to determine the cocoon breakout time $t_{\rm bo,c}$ and the Lorentz factor of the cocoon at breakout $\gamma_{\rm bo,c}$.
 With these values, we can then calculate the breakout energy radiated using Eq.~\eqref{eq:E_bo} and compare it with the measured isotropic energy.
    
    To account for the uncertainties in the measurements, we consider a sample of values for $t_{\rm bo, obs}$ and $E_{\rm pk}$ assuming that they are normally distributed around $t_{\rm GRB}$ and $E_{\rm pk,GRB}$ with standard deviations $\sigma_{t_{\rm GRB}}$ and $\sigma_{E_{\rm pk,GRB}}$, respectively:
\begin{equation}
    t_{\rm bo, obs} \sim \mathcal{N}(t_{\rm GRB}, \sigma_{t_{\rm GRB}})
\end{equation}
and
\begin{equation}
    E_{\rm pk} \sim \mathcal{N}(E_{\rm pk,GRB}, \sigma_{E_{\rm pk,GRB}})
\end{equation}
	Then, for each pair of ($t_{\rm bo, obs}$, $E_{\rm pk}$) values sampled, we calculate the radiated breakout emission, $E_{\rm bo}=E_{\rm bo}(t_{\rm bo,obs}, E_{\rm pk})$, and define a likelihood function as
\begin{equation}
   p(E|t_{\rm bo, obs}, E_{\rm pk}) \propto \exp \left[ \left( \frac{E-E_{\rm bo}}{\sqrt{2}\sigma_{E_{\rm GRB}}} \right)^2 \right],
\end{equation}
where $\sigma_{E_{\rm GRB}}=10^{46}~{\rm erg}$.
	We finally calculate the evidence for each ejecta model as
\begin{eqnarray}
    p(E) = \int p(E|t_{\rm bo, obs}, E_{\rm pk}) \nonumber \\ \times
    p(E_{\rm pk}) p(t_{\rm bo, obs}) dE_{\rm pk} dt_{\rm bo,obs}.
    \label{eq:evidence1}
\end{eqnarray}

\begin{figure}
    \centering
    \includegraphics[width=0.85\linewidth]{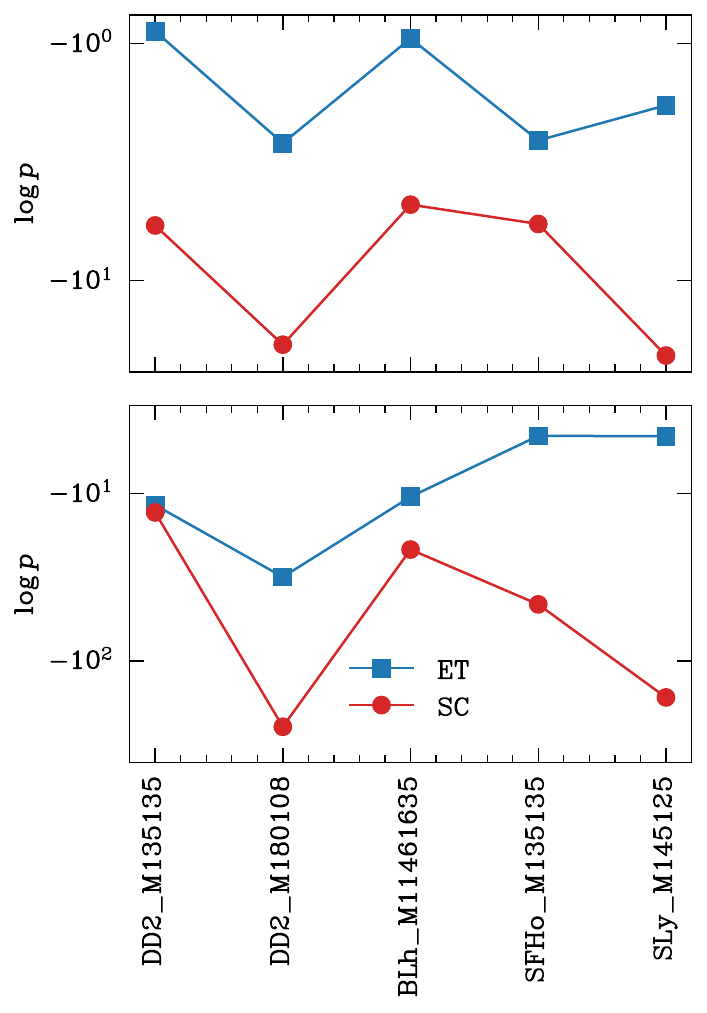}
    \caption{Log likelihood for all five ejecta profiles. The top panel corresponds to the analysis discussed in Sec. \ref{subsec:bayesian_I}, where the isotropic energy radiated is calculated as a function of the observed breakout time and the Lorentz factor of the emitting plasma, which are derived from GRB 170817A.
    The bottom panel corresponds to the analysis discussed in Sec. \ref{subsec:bayesian_II}, where the three observables $(t_{\rm bo, obs}, E_{\rm bo}, E_{\rm pk})$ are calculated from the jet and cocoon propagation simulations considering uniformly distributed engine parameters and then compared to those derived from GRB 170817A.
    In both cases, the blue (red) curve with squares (circles) shows the case of ejecta profiles with an extended tail (sharp cutoff).}
    \label{fig:logp}
\end{figure}

	In the top panel of Fig.~\ref{fig:logp}, we show the logarithm of the evidence evaluated from Eq.~\eqref{eq:evidence1} for the parameters of GRB 170817A.
	The blue (red) curve with squares (circles) corresponds to the ejecta profiles with an extended tail (sharp cutoff).
    It can be seen that the profiles with an extended tail are highly favored for all the simulations, presenting small differences among them.
	On the contrary, the profiles with a sharp cutoff present much larger differences across the different NR simulations.
	This happens because they largely overestimate the radiated energy by $1$-$2$ orders of magnitude due to the large mass in the outermost ejecta layers.
 In Appendix~\ref{ap:ap_bayes}, we show the Bayes factors $BF_{\rm AB}=p_{\rm A}(y)/p_{\rm B}(y)$ across the different simulations we have considered in this study.

\subsection{Jet engine parameter estimation}
\label{subsec:bayesian_II}

	In this subsection, we estimate the likelihood of each ejecta profile following a different approach which also takes into account our propagation model.
	Instead of deriving $t_{\rm bo,c}$ and $\gamma_{\rm bo}$ from the GRB 170817A observables, we take the values resulting from our jet propagation simulations described in Sec.~\ref{sec:results}, and then derive independently the three observables ($t_{\rm bo, obs}$, $E_{\rm bo}$, $E_{\rm pk}$) from them.
	With these observables, we apply a Bayesian analysis equivalent to the one performed in the previous section, but now to also estimate the most likely engine parameters $(B, t_{\rm del})$.
 
	For the given set of observables $y=(t_{\rm GRB}, E_{\rm GRB}, E_{\rm pk,GRB})$, we estimate the posterior distribution for the model parameters $B$ and $t_{\rm del}$ to be
	\begin{equation}
		p(B,t_{\rm del}|y) \propto p(y|B, t_{\rm del}) p(B) p(t_{\rm del}),
	\end{equation}
where we assume uniformly distributed priors:
\begin{equation}
    B\sim \mathcal{U}(B_{\rm min}, B_{\rm max}),~~~t_{\rm del} \sim \mathcal{U}(t_{\rm min}, t_{\rm max}),
\end{equation}
where $B_{\rm min}=5\times 10^{14}~{\rm G}$, $B_{\rm max}=10^{16}~{\rm G}$, $t_{\rm min}=0.1~{\rm s}$, $t_{\rm max}=2~{\rm s}$,
and the likelihood is
\begin{multline}
	p(y|B, t_{\rm del}) \propto \prod_\alpha \exp \left[-\left( \frac{ y_\alpha-Y_\alpha (t_{\rm del}, B)}{\sigma_{\rm t_{\rm GRB}}^2} \right)^2 \right].
\end{multline}
Here, $Y_\alpha(t_{\rm del}, B) = \{ t_{\rm bo, obs}, E_{\rm bo}, E_{\rm pk} \}$, where $\alpha=\{1,2,3\}$ are the predicted observables.	
	
\begin{figure*}
    \centering
    \includegraphics[width=0.995\textwidth]{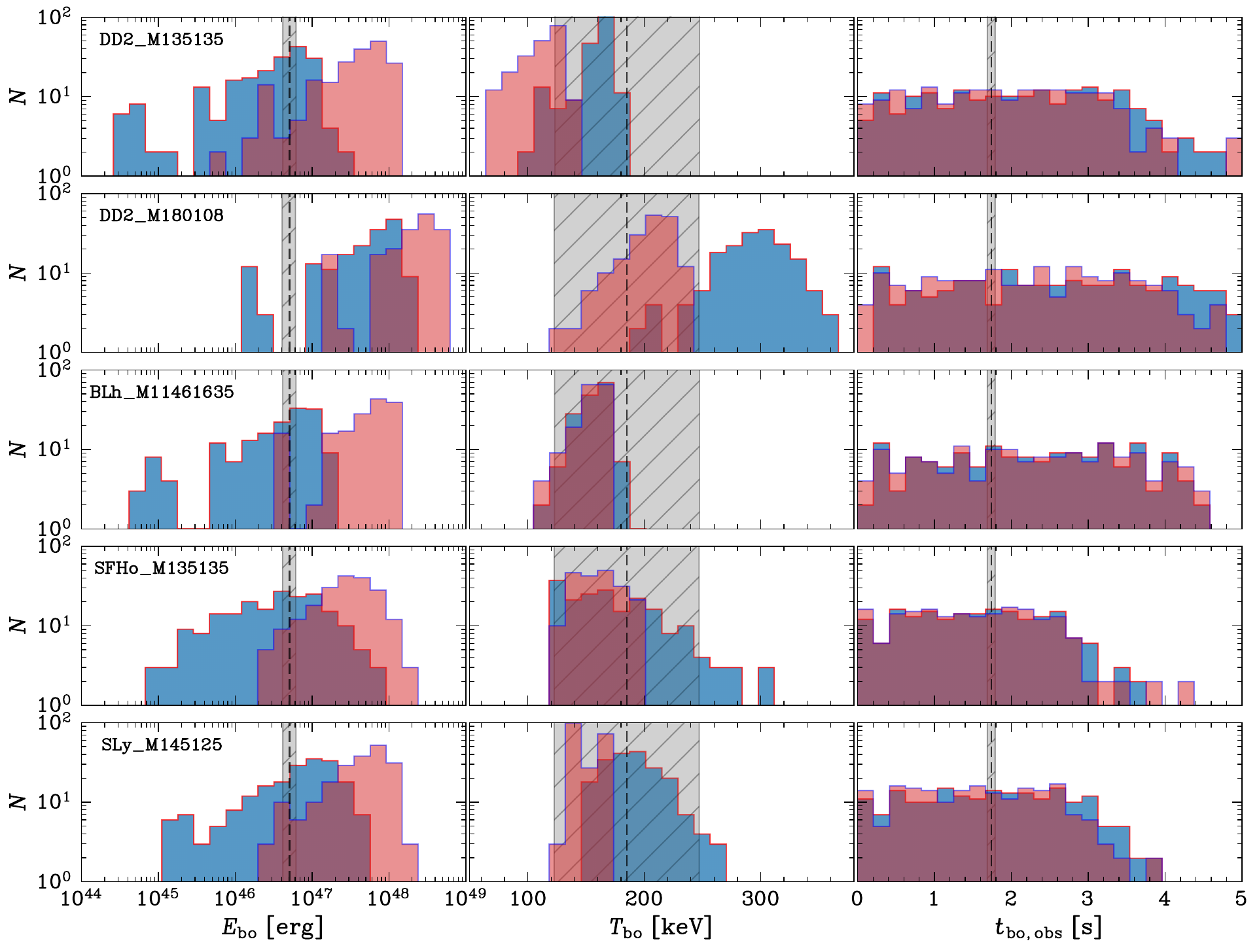}
    \caption{The histograms of the three main observables derived from the cocoon breakout emission are shown. Left column: isotropic energy radiated at breakout. {\it Middle-column panels:} observed temperature at breakout. Right column: observed breakout time. Each row corresponds to a given simulation considered to model the ejecta profile. In each panel, the bars in blue account for simulations where the ejecta profile has an extended tail, while those in red account for simulations where the ejecta profile has an abrupt cutoff at $r=r_{\rm max}(t)$. The vertical dashed line corresponds to the data from GRB 170817A and the gray-filled area denotes their respective uncertainty.}
    \label{fig:histogram}
\end{figure*}

	We first generate histograms of these three observables for each of our five ejecta simulations and the two modeling approaches for the outermost ejecta layers, as depicted in Fig.~\ref{fig:histogram}.
	The blue (red) bars correspond to ejecta profiles with an extended tail (sharp cutoff), the vertical dashed line shows the observables for GRB 170817A and the gray-filled area shows their respective uncertainties.
	In the left panel, we can see that breakouts of ejecta profiles with a sharp cutoff systematically radiate more energy than those of extended tails.
	This is because, as we anticipated, the breakout occurs at $R_{\rm bo,c} \approx r_{\rm max}(t_{\rm bo,c})$ in these scenarios, where the ejecta density is still very high.
	In contrast, ejecta profiles with an extended tail have $R_{\rm bo,c} > r_{\rm max}(t_{\rm bo,c})$, and the densities involved are much lower.
	Moreover, the isotropic energy output during breakout from an ejecta outer boundary with a sharp cutoff is larger despite the Lorentz factor of the emitting material being systematically lower than that for a breakout with an extended ejecta tail, as can be noticed in the middle panels of Fig. \ref{fig:histogram}.

	When comparing the different ejecta simulations (from top to bottom), we can notice that the breakout signal from DD2\_M180-108 shows the highest observed temperature for both models of the ejecta outer layers, due to the high Lorentz factors of the cocoon at the breakout (see also Fig. \ref{fig:cocoon_Gamma}).
	The main reason is that the ejecta derived from this simulation is the most massive one and this slows down the cocoon expansion.
    As a result, the cocoon is more compact and exerts a larger pressure on the jet, which is then easily collimated.
	The larger mass in this case also delays the breakout giving the shock more time to accelerate.
    However, this delay is not reflected in the observed breakout time since the breakout radius is also larger [see Eq. \eqref{eq:t_bo_obs}].
	The larger breakout radii, ejecta mass, and Lorentz factors imply that this simulation also shows the largest isotropic energy radiated.
	On the contrary, DD2\_M135-135 has the least massive ejecta, which explains the lower overall temperatures observed at breakout.

 Finally, the observed breakout time $t_{\rm bo,obs}$ covers a wide range for all simulations, though it tends to be smaller for the two cases with a short-lived remnant and a BH engine.
 This is due to two reasons that can make the jet propagate faster: a lower ejecta mass at smaller radii, or a larger jet power for a given magnetic field strength at the engine.

\begin{figure*}
    \centering
    \includegraphics[width=0.995\linewidth]{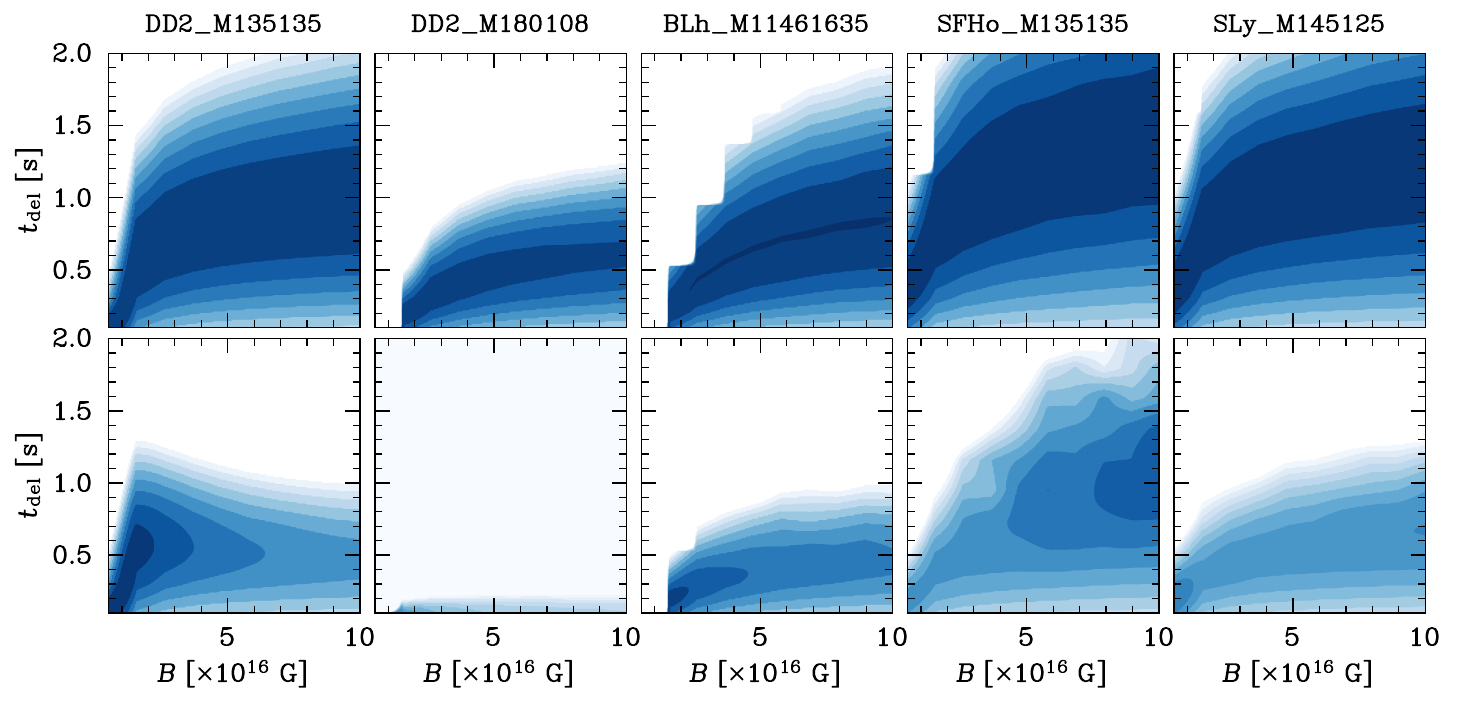}
    \caption{Posterior distributions for the engine parameters $(B, t_{\rm del})$ are shown for each simulation from left to right. The top row corresponds to ejecta profiles with an exponential tail and the bottom row corresponds to profiles with an abrupt cutoff. Each color map is normalized to its maximum and shown with a logarithmic scale.}
    \label{fig:parameter_estimation}
\end{figure*}

    We show the posterior distributions for each simulation and for both models of the ejecta profile in Fig.~\ref{fig:parameter_estimation}.
    The strongest constraint is imposed by the time delay of the GRB arrival relative to GW emission, as it has the lowest relative standard deviation among the three observables.
    This imposes strong restrictions on the allowed values for $t_{\rm del}$, although these are mediated by the magnetic field of the engine $B$ and the ejecta profile.
    As a general rule, ejecta profiles with a lower total mass allow longer time delays since in these environments the jet propagates faster.
    Larger time delays are also favored for BH engines compared to NS engines, since the jet is more powerful in the first case for the same magnetic field value.
    
    Ejecta profiles with an extended tail are largely favored in our analysis; ejecta profiles with a sharp cutoff tend to produce breakout signals that are significantly more energetic than GRB 170817A.
    This is especially noticeable with DD2\_M180-108, which in addition involves significantly larger breakout speeds and, therefore, largely overestimates the peak energy for the breakout emission.

	To quantify the likelihood of each model, we calculate the evidence for each NR simulation as
\begin{multline}
	p(y) = \int dt_{\rm del}dB p(t_{\rm del}) p(B) p(y|B, t_{\rm del};\mathcal{S}) \simeq \\
	\frac{1}{2 N_{t_{\rm del}} N_{B}}\sum_{\theta_{\rm j}, t_{\rm del}, B} p(y|B,t_{\rm del};\mathcal{S}),
\end{multline}
where $N_{t_{\rm del}}$ and $N_{B}$ are the number of data points sampled for the jet launching delay and the dipolar magnetic field at the engine, respectively, and the factor of $2$ arises from the fact that we sum the likelihood for the two opening angles considered in our analysis.

Finally, in the lower panel of Fig. \ref{fig:logp}, we show the logarithmic likelihood for the five NR simulations and the two modeling approaches for the outermost boundaries of the ejecta.
The blue (red) curve with squares (circles) corresponds to an ejecta profile with an extended tail (sharp
cutoff).
As in the analysis from the previous section, ejecta profiles with an extended tail are favored for all the simulations over profiles with a sharp cutoff.
In both cases, DD2\_M180-108 ranks the worst due to its overestimation of the isotropic energy radiated (for profiles with a sharp cutoff) or the temperature (for profiles with an extended tail).
The highest likelihood is reached by the simulations with a BH engine (provided that the profile has an extended tail).
The Bayes factors $BF_{\rm AB}=p_{\rm A}(y)/p_{\rm B}(y)$ across the different simulations and ejecta profiles with this approach are also shown in the Appendix \ref{ap:ap_bayes}.

\section{Discussion} \label{sec:discussion}

    We have developed a model to calculate the electromagnetic signal from relativistic breakouts of expanding ejecta in the context of NS mergers, using realistic ejecta profiles derived from NR simulations.
	The model involves three main physical components: the ejecta, the engine, and the jet+cocoon; and the evolution was divided into three phases: the propagation of the jet and the cocoon through the ejecta, the cocoon propagation after the jet head has broken out, and the evolution of the cocoon after it breaks out.

    We first derived the rest-mass density and velocity profiles of the ejecta from five long-term NS merger simulations, which comprise three different mass ratios and four different nuclear EOS.
	To derive these profiles, we followed two different approaches when accounting for the outermost layers of the ejecta: we either considered that the ejecta has a sharp cutoff at a radius $r_{\rm max}(t) \approx \beta_{\rm ej,max} c t$ or fast-moving tail extending for $r > r_{\rm max}(t)$.
    Such an extended tail may be produced by small perturbations in the velocity of the outflow, due for example to thermal motion, or by the interaction between different ejecta layers through pressure gradients.

	The second component of our model is the engine that powers the jet and the cocoon.
	The NR simulations we derived the ejecta profiles from did not include magnetic fields, which are a key ingredient for jet launching.
	Instead, we have analytically modeled the jet power for the two possible types of central engine, a BH or a magnetar, and collected the uncertainties into three free parameters: the dipolar magnetic field threading the engine (the NS itself in the magnetar case and the event horizon in the BH case) $B$, the time delay between the merger and the jet launch $t_{\rm del}$, and the initial opening angle of the jet $\theta_{\rm j,0}$.
    In reality, the jet may be launched at $t<t_{\rm del}$ but we assume that only becomes powerful and fast enough for $t\gtrsim t_{\rm del}$.

	The final component of our model is the jet+cocoon system.
	We have solved a system of equations for the evolution of this system by separating the propagation in three different phases.
    First, for $t_{\rm del} < t < t_{\rm bo,h}$, the jet is launched by the engine and propagates through the ejecta.
    During this phase, the jet head energizes the hot cocoon.
    For $t_{\rm bo, h} < t < t_{\rm bo,c}$, the jet head has already broken out of the ejecta and detaches from the cocoon; the latter, on the contrary, is still within the ejecta and propagates out as a hot adiabatic fireball.
    In reality, the cocoon and the jet head will form a continuous angle-dependent structure, and the breakout will occur at different times for different angles.
    Given that the cocoon is expected to be only mildly relativistic, we have averaged its properties for different polar angles to model it as a single structure that behaves as (a portion of) a spherical blast wave.
    Finally, after the forward shock driven by the cocoon breaks out ($t \gtrsim t_{\rm bo, c}$), the energized (shocked) ejecta layers expand adiabatically until they become transparent to radiate out their internal energy.


Most previous numerical and analytical works on jet propagation in NS merger environments assumed analytical ejecta profiles with a sharp cutoff at the time-dependent outermost radius $r_{\rm max}(t)\simeq v_{\rm max} t$ \cite{Nagakura_etal2014ApJ...784L..28N, Gottlieb_etal2018a, Gottlieb_etal2018b, Salafia_etal2020, Hamidani_etal2020, HamidaniIoka2021, Gottlieb_etal2022}.
    Although this does not significantly affect the time and radius of breakout, it does have a large influence on the emitted shock breakout signal \cite{Nakar2020}.
    The primary effect is that the energy radiated is smaller when the ejecta profile has an extended tail similar to the scenario explored by
    Ref.~\cite{BeloborodovLundman2020}, where they proposed that NS mergers eject an ultrarelativistic low-mass envelope such that the breakout occurs before the shock reaches the outermost layers of the ejecta (as in our extended tail scenario).
    
Our shock breakout emission model builds upon the scenario investigated in Ref. \cite{Gottlieb_etal2018b} (see also Ref. \cite{Gottlieb_etal2018a}), which studied jet propagation and cocoon breakout through numerical simulations.
This work assumed an analytically prescribed ejecta profile with a power-law density distribution featuring a sharp cutoff.
Since the jet evolution is treated numerically, the merger ejecta may naturally develop an extended outer tail, similar to the profiles we consider.
As we have demonstrated, this extended tail plays a crucial role in shaping the shock breakout emission.

Other recent studies have also explored jet propagation, cocoon breakout, and its associated emission~\cite{HamidaniIoka2023a, HamidaniIoka2023b}.
However, these works primarily focused on the cooling emission which occurs over longer timescales and at lower energies rather than the prompt breakout signal itself.
Additionally, early analytical treatments of shock breakout in relativistic outflows~\cite{NakarSari2012}, provided general scaling relations but only in a static ambient medium, such as the case of a jet propagating through a star envelope.

Our model presents several key improvements: (i) Unlike previous approaches, which often treated jet$+$cocoon propagation and breakout emission separately, we have developed a semianalytical framework that self-consistently tracks these various stages.
This provides a more comprehensive understanding of the full process connecting the original binary NS properties with the shock breakout signal.
(ii) We have used realistic ejecta profiles extracted from NR simulations of binary neutron star mergers.
In addition, we have explored in detail the impact of the outermost ejecta layers on the shock breakout signal by considering scenarios where the ejecta profile has a sharp cutoff or an extended tail.
We demonstrated that an extended low-density tail is essential to reproduce the observed energetics of GRB 170817A, while models with a sharp cutoff significantly overestimate the breakout luminosity.
(iii) By incorporating a Bayesian analysis, we systematically compare our model predictions with GRB 170817A data, exploring different ejecta structures, binary configurations, and NS properties.
This enables us to infer constraints on the central engine and merger parameters, providing a direct link between $\gamma$-ray observations and the merger dynamics.
In particular, we find that the NR simulation DD2\_M180-108, which corresponds to a high mass ratio ($q \sim 1.7$), is strongly disfavored, whereas models with softer equations of state and prompt BH formation are slightly preferred.

These results align well with previous constraints on the NS EOS derived from GW170817 and AT2017gfo \cite{Breschi_etal2021MNRAS.505.1661B}.
Future detections of similar events will further refine these constraints and provide new insights into the physics of relativistic outflows and electromagnetic counterparts to compact object mergers.

\section{Conclusions} \label{sec:conclusions}

    We have investigated the propagation of relativistic jets with their associated cocoons through the ejecta from neutron star mergers; the ejecta profiles were derived from general-relativistic radiation hydrodynamic simulations.
    We have calculated the resulting cocoon shock breakout emission by modeling the shape of the outer layers of the ejecta along with the central engine parameters to compare our predictions with those of the sGRB associated with GW170817.

    We found that the properties of the outermost ejecta layers significantly impact the shock breakout emission.
    The transition from NS merger ejecta to the interstellar medium can feature either a sharp density cutoff or a smooth, extended fast-moving tail.
    Jets moving through ejecta with a smooth extended tail break out at later times and larger radii, where the mass density of the ejecta is significantly lower.
    This allows higher Lorentz factors for the jet head and the cocoon compared to the case where the ejecta profile has a sharp cutoff.
    Because of the lower densities involved, ejecta profiles with an extended tail result in less energetic cocoon breakout emission, which aligns well with the observed properties of GRB 170817A. In contrast, ejecta profiles with a sharp cutoff tend to overestimate the energy radiated at breakout.

    We performed a Bayesian analysis to estimate the engine parameters that can better explain the observed data from GRB 170817A.
    We found that scenarios that lead to an early BH formation are slightly favored over the scenarios explored that have a long-lived magnetar engine.
    In particular, we found the simulation with DD2 NS EOS and a large binary mass ratio ($q=1.67$) to be the least consistent with the observed data, since it tends to overestimate both the total energy radiated at the breakout and the peak energy of the spectrum.

    Our model has some limitations.
The exact shape of the extended ejecta tail is somewhat uncertain, as it was modeled based only on the velocity dispersion of the outermost ejecta layers due to thermal motion; namely, we did not take into account other effects such as pressure gradients that may spread the outermost layers even more.
Additionally, our analysis assumes a uniform cocoon breakout, while a more realistic scenario would involve a stratified breakout with the jet head breaking out of the outer layers of the ejecta first, followed by other portions of the expanding cocoon material.
Finally, some of our conclusions are based on a single observed event, limiting the strength of our Bayesian analysis.

In future work, we will refine our model using detailed general-relativistic magnetohydrodynamics simulations to better understand the properties of the relativistic jet at its launch time. We will also further explore the impact of angle-dependent breakout for the cocoon on the resulting light curves.
We will also extend our study to a larger set of NS equations of state and binary mass ratios.

\vspace{0.5cm}

{\it Note added.} While this manuscript was under review, \citet{Rosswog_etal2024} presented a similar but complementary analysis of the effects of the fast ejecta on electromagnetic observables by using Lagrangian simulations. Our results on the shock breakout emission broadly agree with each other.

\begin{acknowledgments}

We thank the anonymous referee for useful comments that helped to improve the overall quality of the paper.
We thank Om Sharan Salafia, Geoffrey Ryan, Christopher Irwin, Ehud Nakar, and Kunihito Ioka for insightful discussions and comments.
E. M. G., K. M., and D. R. acknowledge support from the National Science Foundation under Grant No.~AST-2108467. 
M. B. acknowledges support from the Eberly Research Fellowship
at the Pennsylvania State University and the Simons
Collaboration on Extreme Electrodynamics of Compact
Sources (SCEECS) Postdoctoral Fellowship at the Wisconsin
IceCube Particle Astrophysics Center (WIPAC), University
of Wisconsin-Madison. 
D. R. acknowledges support through an Alfred P.~Sloan Foundation Fellowship and funding from the U.S. Department of Energy, Office of Science, Division of Nuclear Physics under Awards No. DE-SC0021177 and No. DE-SC0024388 and from the National Science Foundation under Grants No. PHY-2011725, No. PHY-2020275, and No. PHY-2116686.
S. B. acknowledges support by the EU Horizon under ERC Consolidator Grant, No. InspiReM-101043372.

Part of the simulations were performed on SuperMUC-NG at the Leibniz-Rechenzentrum (LRZ) Munich. The authors acknowledge the Gauss Centre for Supercomputing e.V. (\url{www.gauss-centre.eu}) for funding this project by providing computing time on the GCS Supercomputer SuperMUC-NG at LRZ (allocations {\tt pn68wi}, {\tt pn36jo} and {\tt pn39go}). Part of the numerical simulations were performed on the national HPE Apollo Hawk at the High-Performance Computing Center Stuttgart (HLRS). The authors acknowledge HLRS for funding this project by providing access to the supercomputer HPE Apollo Hawk under the Grant No. INTRHYGUE/44215 and No. MAGNETIST/44288.
Simulations were also performed on TACC's Frontera (NSF LRAC allocation PHY23001), on NERSC's Perlmutter, and on the Pennsylvania State University’s Institute for Computational and Data Sciences’ Roar supercomputer. This research used resources of the National Energy Research Scientific Computing Center, a DOE Office of Science User Facility supported by the Office of Science of the U.S.~Department of Energy under Contract No.~DE-AC02-05CH11231.

\end{acknowledgments}

\section*{Data availability}

The data that supports the findings of this article are not publicly available.
The data are available from the authors upon reasonable request.

\appendix

\section{Ejecta modeling}
\label{ap:ejecta_modeling}

The outflow rest-mass density at a time $t$ and radius $r$ is composed of matter that has crossed the sphere of radius $R_0 \approx 442~{\rm km}$ at a previous time $t_0$ with an asymptotic velocity $v_\infty(t_0) \approx (r-R_0)/(t-t_0)$.
Since several portions of the ejecta may satisfy the above condition, we weigh the relevance of each of them at $(r, t)$ using a function $\mathcal{F}[\Delta v]$, where $\Delta v := v_\infty(t_0) - (r-R_0)/(t-t_0)$, such that it has a maximum at $\Delta v=0$, and decays smoothly in both directions.
	We choose a normal distribution $\mathcal{F}(\Delta v) = (2\pi \sigma^2)^{-1/2} e^{-(\Delta v)^2/2\sigma^2}$ based on the fact that the particles in the outflow are expected to follow a Maxwell--Boltzmann distribution with standard deviation $ \sigma \approx \sqrt{k_{\rm B}T/m_{\rm p}} \approx 0.025 c$ for a typical ejecta temperature $T \sim 5\times 10^9~{\rm K}$.

	If time $t$ exceeds the duration of the simulation, we extrapolate the data assuming that the ejecta mass-loss rate and velocity distribution follow a power law with the same slope as the one measured at late times.
Finally, taking into account the rate of mass crossing the radius $r$, $\dot{M}(r)=4\pi r^2 \Gamma v_{\rm ej} \rho'_{\rm ej}(r,t)$, we calculate the rest-mass density profile by integrating over time as
	\begin{equation}
		\rho_{\rm ej}(r, t) \simeq \frac{1}{4\pi r^2} \int_{0}^t  \frac{dt_0}{\Gamma_\infty (t-t_0)} \dot{M}(t_0)\mathcal{F}( \Delta v ).
		\label{eq:rho_r}
	\end{equation}

	Similarly, to determine a unique velocity profile $v_{\rm ej}(r, t)$, we integrate in time the asymptotic velocity measured at $R_0$ weighted by the mass-loss rate,
\begin{equation}
	v_{\rm ej}(r, t) \simeq \frac{1}{4\pi r^2 \rho_{\rm ej}(r,t)} \int_{0}^t  \frac{v_\infty (t_0) dt_0}{\Gamma_\infty (t-t_0)} \dot{M}(t_0)\mathcal{F}( \Delta v ).
	\label{eq:v_r}
\end{equation}

\section{Jet and Cocoon propagation}
\label{ap:jet_propagation}

Here, we derive the equations governing the evolution of the jet and cocoon through the expanding ejecta derived from our numerical relativity simulations.

\subsection{Jet head evolution}
    While the bulk of the jet moves with a relativistic Lorentz factor $\Gamma_{\rm j} \gg 1$, at the head, it must decelerate to match the mildly relativistic ejecta speed.
    This will likely occur through the formation of a forward/reverse shock structure; however, in the case of a magnetically dominated jet, the reverse shock may be weak or absent such that the deceleration occurs more smoothly.
    The unshocked jet can be described by an energy-momentum tensor given by
\begin{equation}
    T_{\rm j}^{\mu \nu} = (w_{\rm j} + b_{\rm j}^2) u_{\rm j}^\mu u_{\rm j}^\nu + (p_{\rm j} + b_{\rm j}^2/2) \eta^{\mu \nu} - b_{\rm j}^\mu b_{\rm j}^\nu,
\end{equation}
where $w_{\rm j}$ is the specific enthalpy, $p_{\rm j}$ is the thermal pressure, $u_{\rm j}^\mu \simeq \Gamma_{\rm j} c( 1, 0, 0, \beta_{\rm j})$ is the jet $4$-velocity expressed in cylindrical coordinates, $b_{\rm j}^\mu$ is the magnetic field 4-vector in the jet frame, and $\eta^{\mu\nu}$ is the inverse spacetime metric.
    Since at large distances from the engine, the toroidal component of the magnetic field $B^\phi$ in the lab frame is much larger than the poloidal component $B$, we can neglect the latter and express $b_{\rm j}^\mu = (0, 0, b_{\rm j}, 0)$, where $b_{\rm j}= B_{\rm j}^\phi / (\sqrt{4\pi} \Gamma_{\rm j})$ is the proper magnetic field.
    The jet luminosity is obtained by integrating the ``$0z$'' component of the energy-momentum tensor on a cross section of the jet,
\begin{equation}
    L_{\rm j} = \int_0^{r_{\rm j}} T_{\rm j}^{0z} 2\pi r dr \simeq
                \Sigma_{\rm j} \Gamma_{\rm j}^2 \beta_{\rm j} c (w_{\rm j} + b_{\rm j}^2),
    \label{eq:jet_lum}
\end{equation}
where $\Sigma_{\rm j}:= \pi r_{\rm j}^2$ is the jet cross section and $r_{\rm j}$ is the cylindrical radius of the jet.

\begin{figure}
    \centering
    \includegraphics[width=0.645\linewidth]{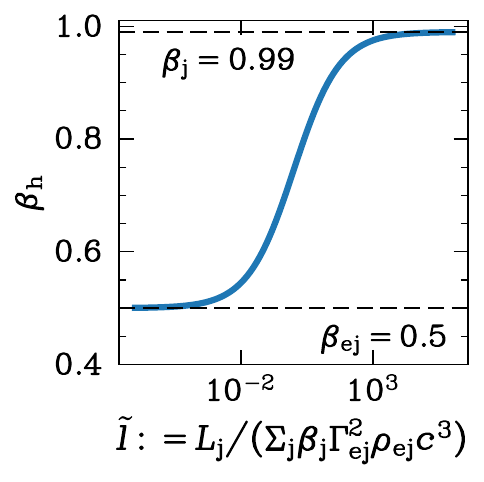}
    \caption{Jet head velocity $\beta_{\rm h}$ as a function of the ratio of the energy density between the jet and the ejecta, $\tilde{l}$, for the values of jet velocity $\beta_{\rm j}=0.99$ and ejecta velocity $\beta_{\rm ej}=0.5$.}
    \label{fig:betas}
\end{figure}

    Momentum balance across the forward and reverse shock regions gives
\begin{equation}
    (w_{\rm j} + b_{\rm j}^2) \beta_{\rm jh}^2 \Gamma_{\rm jh}^2 + p_{\rm j} + b_{\rm j}^2/2 = w_{\rm ej} \beta_{\rm h,ej}^2 \Gamma_{\rm h,ej}^2 + p_{\rm ej},
    \label{eq:p_balance}
\end{equation}
where $\Gamma_{\rm jh} = \Gamma_{\rm j} \Gamma_{\rm h} ( 1 - \beta_{\rm j} \beta_{\rm h})$ is the relative Lorentz factor between the jet and the head and $\beta_{\rm jh}:= (\beta_{\rm j} - \beta_{\rm h})/(1-\beta_{\rm j}\beta_{\rm h})$ is its relative velocity; a similar definition applies for the ejecta quantities on the right-hand side of Eq.~\eqref{eq:p_balance}.
    We assume that the ejecta is sufficiently cold and take $p_{\rm ej} \approx 0$ and $w_{\rm ej} \approx \rho_{\rm ej} c^2$.
    Equation~\eqref{eq:p_balance} then reduces to
\begin{equation}
    \tilde{l}~ \Gamma_{\rm h}^2 (\beta_{\rm j}-\beta_{\rm h})^2 + \tilde{p} = \Gamma_{\rm h}^2 (\beta_{\rm h} - \beta_{\rm ej})^2,
    \label{eq:p_balance_2}
\end{equation}
where
\begin{equation}
    \tilde{l} := \frac{L_{\rm j}}{\Sigma_{\rm j} \beta_{\rm j} \Gamma_{\rm ej}^2 \rho_{\rm ej} c^3}, ~~~{\rm and} ~\tilde{p}:= \frac{p_{\rm j}+ b_{\rm j}^2/2}{\Gamma_{\rm ej}^2 \rho_{\rm ej} c^2}
    \label{eq:l_p}
\end{equation}
are the ratio of the energy density between the jet and the ambient matter and the ratio between the jet pressure and the energy density of the ambient matter, respectively.
    The solution for the jet head velocity from Eq. \eqref{eq:p_balance_2} is
\begin{equation}
    \beta_{\rm h} = \frac{\tilde{l}\beta_{\rm j}-\beta_{\rm ej}-\left[ (\tilde{l}\beta_{\rm j} - \beta_{\rm ej})^2 - ( \tilde{l} \beta_{\rm j}^2 + \tilde{p}-\beta_{\rm ej}^2)(\tilde{l}-\tilde{p}-1) \right]^{1/2}}{\tilde{l} - \tilde{p} - 1}.
    \label{eq:beta_h}
\end{equation}
    For a classical cold hydrodynamic jet, $\tilde{p}$ can be neglected and Eq. \eqref{eq:beta_h} reduces to the simpler expression \cite{Salafia_etal2020, HamidaniIoka2021}
\begin{equation}
    \beta_{\rm h} = \beta_{\rm ej} + \frac{\beta_{\rm j} - \beta_{\rm ej}}{1 + \tilde{l}^{-1/2}}.
    \label{eq:beta_h_2}
\end{equation}
    It is worth noting that even for a large magnetization where $b_{\rm j}^2 \gg w_{\rm j}$, from Eqs. \eqref{eq:jet_lum} and the definitions of Eq. \eqref{eq:l_p}, we get $\tilde{p} \simeq \tilde{l}/(2\Gamma_{\rm j}^2) \ll \tilde{l}$, and the expression for the head velocity also simplifies to Eq. \eqref{eq:beta_h_2}.
    Since $\beta_{\rm j} \approx 1$, the head velocity depends primarily on the ratio of the jet energy density to the ejecta rest-mass energy density as well as on the ejecta speed $\beta_{\rm ej}$ in that position.
    In Fig. \ref{fig:betas}, we show an example of how $\beta_{\rm h}$ depends on $\tilde{l}$ for $\beta_{\rm j}=0.99$ and $\beta_{\rm ej}=0.5$.

\subsection{Cocoon evolution}

    The cocoon is formed by the heated matter that enters the jet head through the forward shock and is pushed sideways.
    The energy per unit of time deposited in the cocoon by the jet is
\begin{equation}
    \frac{dE_{\rm c}}{dt} = \eta L_{\rm j}(t_{\rm eng}) \left[ \beta_{\rm j} - \beta_{\rm h} \right],
\end{equation}
where $t_{\rm eng}:= t - (z_{\rm h}-z_0)/c$ is the retarded time at the engine, and $\eta$ accounts for the fraction of the jet head that is in causal contact with the cocoon and can be parametrized as \cite{Salafia_etal2020}
\begin{align}
    \eta =
    \begin{cases}
        \frac{2}{\mu} - \frac{1}{\mu^2}, ~~~ \mu > 1, \\
        1, ~~~ \mu \leq 1,
    \end{cases} 
    \label{eq:eta}
\end{align}
where $\mu = \sqrt{3} \Gamma_{\rm h} \beta_{\rm h} \theta_{\rm j}$.

    We assume that the cocoon energy density is dominated by radiation and that it is distributed uniformly throughout its volume.
    We consider that the cocoon pressure takes a time $\Delta t_{\rm c} \sim (z_{\rm h} - z_0)/2c_{\rm s}$ to build up, where $c_{\rm s} \approx c/\sqrt{3}$ is the sound speed.
    Then, the cocoon pressure at time $t$ can be calculated as
\begin{equation}
    p_{\rm c} = \frac{E_{\rm c}(t-\Delta t_{\rm c})}{3V_{\rm c}},
    \label{eq:p_c}
\end{equation} 
where $V_{\rm c} = \frac{2\pi}{3} (z_{\rm h}-z_0) r_{\rm c,\perp}^2$ is the volume of an ellipsoid of radius $r_{\rm c}$ and height $z_{\rm h}-z_0$.

    The cocoon expands laterally at a speed
\begin{equation}
    \beta_{\rm c,\perp} \approx \left( 1 + \bar{\rho}_{\rm ej} c^2/p_{\rm c} \right)^{-1/2} + \beta_{\rm ej, \perp},
    \label{eq:beta_c}
\end{equation}
where $\bar{\rho}_{\rm ej}$ is the rest-mass density of the ejecta averaged over the cocoon volume.
    The first term in Eq. \eqref{eq:beta_c} accounts for the balance between the cocoon pressure and the ram pressure of the ejecta whereas the second term accounts for the lateral expansion of the ejecta itself, which can be calculated as
\begin{equation}
    \beta_{\rm ej, \perp} \approx \beta_{\rm ej}(d_*) \left( \frac{r_{\rm c,\perp}}{d_*} \right),
    \label{eq:beta_a_perp}
\end{equation}
where $d_*=\sqrt{r_{\rm c,\perp}^2 + \left[\frac{1}{2}(z_{\rm h}+z_0)\right]^2}$ is the distance to the contact discontinuity between the cocoon and the ejecta at half the height of the cocoon.

\subsection{Jet collimation} \label{ssec:collimation}

    The pressure exerted by the cocoon can collimate the jet provided it is higher than the jet thermal pressure $p_{\rm j}(z) \sim L_{\rm j} z_0^2 / (4\pi \theta_{\rm j,0}^2 z^4 c)$  \cite{Matzner2003, Salafia_etal2020}.
    If this is the case, a reconfinement shock forms \cite{Morsony_etal2007}.
    The shock surface is determined by the balance between the cocoon pressure and the jet ram pressure normal to the shock.
    In the comoving frame of the ejecta, the pressure balance equation reads \cite{Salafia_etal2020}
\begin{equation}
    (w_{\rm j} + b_{\rm j}^2) \Gamma_{\rm j,ej}^2 \beta_{\rm j,ej}^2 \Gamma_{\rm ej}^{-2} \left( \frac{r_{\rm s}}{z-z_0} - \frac{dr_{\rm s}}{dz} \right)^2 + p_{\rm j} + \frac{b_{\rm j}^2}{2} = p_{\rm c},
    \label{eq:reconfinement_shock}
\end{equation}
where $r_{\rm s}$ is the cylindrical radius of the reconfinement shock.
This can be rewritten as
\begin{equation}
    \tilde{l} (\beta_{\rm j} - \beta_{\rm ej})^2 \left( \frac{r_{\rm s}}{z-z_0} - \frac{dr_{\rm s}}{dz} \right)^2 + \frac{\tilde{l}}{2\Gamma_{\rm j}^2} = \tilde{p}.
\end{equation}
  Since $\Gamma_{\rm j} \gg 1$, we can neglect the second term on the left-hand-side; then, Eq. \eqref{eq:reconfinement_shock} has the solution \cite{Bromberg_etal2011}
\begin{equation}
    r_{\rm s}(z) = \theta_{{\rm j},0} ( 1 + A z_{\rm *} ) z - \theta_{{\rm j},0} A z^2,
    \label{eq:r_s}
\end{equation}
where $A = \sqrt{\pi c \beta_{\rm j} p_{\rm c}/L_{\rm j} (\beta_{\rm j}-\beta_{\rm ej})^2}$, and $z_* \approx z_0$ is the height at which the jet internal pressure equals the cocoon pressure.
The reconfinement shock converges to the jet axis, $r_{\rm s}=0$, at a height $\hat{z}= A^{-1}+z_*$.
    We will assume that the jet is conical up to the point where collimation due to the cocoon becomes significant, which is approximately where the reconfinement shock is parallel to the jet axis.
    Since the shock surface given by Eq. \eqref{eq:r_s} has a parabolic shape, we can assume that the jet is collimated for  $z > z_{\rm coll} \simeq (z_* + \hat{z})/2$.
    Above that region, we set the jet cylindrical radius to $\tan (\theta_{{\rm j},0})z_{\rm coll}$.
    It is worth mentioning that $\hat{z}$ varies with time and thus the height at which the jet is collimated and attains a cylindrical radius at the jet head position can vary too.

\subsection{System of equations}

Using Eqs.~\eqref{eq:l_p}, \eqref{eq:beta_h}, \eqref{eq:eta}, \eqref{eq:p_c}, \eqref{eq:beta_c}, and \eqref{eq:beta_a_perp}, together with the ejecta functions given by Eqs.~\eqref{eq:rho_r} and \eqref{eq:v_r} and the collimation criterion, we obtain the following system of coupled first-order ordinary differential equations:
\begin{eqnarray}
    & \dfrac{dz_{\rm h}}{dt} = \beta_{\rm h} c, \nonumber \\
    & \dfrac{dr_{\rm c,\perp}}{dt} = \beta_{\rm c,\perp} c, \nonumber \\
    & \dfrac{dE_{\rm c}}{dt} = \eta L_{\rm j}(t_{\rm eng}) \left[ \beta_{\rm j} - \beta_{\rm h} \right].
\end{eqnarray}

\section{Bayes factors}
\label{ap:ap_bayes}

\begin{table*}
    \centering
    
    \vspace{0.2cm}
    \begin{tabular}{|c|c|c|c|c|c|}
        \hline
         & DD2\_M135-135 & DD2\_M180-108 & BLh\_M1146-1635 & SFHo\_M135-135 & SLy\_M145-125 \\
        \hline
         DD2\_M135-135 & $\dots$  & $1.1$ & $0.1$ & $1.0$ & $0.7$ \\
        \hline
        DD2\_M180-108 & $-1.1$ & $\dots$  & $-1.0$ & $-0.0$ & $-0.4$ \\
        \hline
        BLh\_M1146-1635 & $-0.1$ & $1.0$ & $\dots$  & $1.0$ & $0.6$ \\
        \hline
        SFHo\_M135-135 & $-1.0$ & $0.0$ & $-1.0$ & $\dots$  & $-0.3$ \\
        \hline
        SLy\_M145-125 & $-0.7$ & $0.4$ & $-0.6$ & $0.3$ & $\dots$  \\
        \hline
    \end{tabular}

    \vspace{0.2cm}
    \begin{tabular}{|c|c|c|c|c|c|}
        \hline
         & DD2\_M135-135 & DD2\_M180-108 & BLh\_M1146-1635 & SFHo\_M135-135 & SLy\_M145-125 \\
        \hline
         DD2\_M135-135 & $\dots$  & $16.7$ & $-1.1$ & $-0.1$ & $19.8$ \\
        \hline
        DD2\_M180-108 & $-16.7$ & $\dots$  & $-17.9$ & $-16.8$ & $3.1$ \\
        \hline
        BLh\_M1146-1635 & $1.1$ & $17.9$ & $\dots$  & $1.1$ & $21.0$ \\
        \hline
        SFHo\_M135-135 & $0.1$ & $16.8$ & $-1.1$ & $\dots$  & $19.9$ \\
        \hline
        SLy\_M145-125 & $-19.8$ & $-3.1$ & $-21.0$ & $-19.9$ & $\dots$  \\
        \hline
    \end{tabular}
    
    \vspace{0.2cm}
    \begin{tabular}{|c|c|c|c|c|}
        \hline
         DD2\_M135-135 & DD2\_M180-108 & BLh\_M1146-1635 & SFHo\_M135-135 & SLy\_M145-125 \\
        \hline
        $4.2$ & $19.9$ & $3.0$ & $3.1$ & $23.4$ \\
\hline
    \end{tabular}
    \caption{Logarithms of Bayes factors comparing our models as described in Sec. \ref{subsec:bayesian_I}.
    The top shows different NR simulations for ejecta profiles with an extended tail.
    The middle is the same as the top, but for ejecta profiles with a sharp cutoff.
    The bottom compares the two approaches (extended tail and sharp cutoff) corresponding to the same NR simulation.}

    \label{tab:bayes_1}
\end{table*}

\begin{table*}
    \centering

    \vspace{0.2cm}
    \begin{tabular}{|c|c|c|c|c|c|}
        \hline
         & DD2\_M135-135 & DD2\_M180-108 & BLh\_M1146-1635 & SFHo\_M135-135 & SLy\_M145-125 \\
        \hline
        DD2\_M135-135 & $\dots$ & $27.0$ & $-0.7$ & $-2.9$ & $-2.6$ \\
        \hline
        DD2\_M180-108 & $-27.0$ & $\dots$ & $-27.7$ & $-29.9$ & $-29.6$ \\
        \hline
        BLh\_M1146-1635 & $0.7$ & $27.7$ & $\dots$ & $-2.2$ & $-1.9$ \\
        \hline
        SFHo\_M135-135 & $2.9$ & $29.9$ & $2.2$ & $\dots$ & $0.3$ \\
        \hline
        SLy\_M145-125 & $2.6$ & $29.6$ & $1.9$ & $-0.3$ & $\dots$ \\
        \hline
    \end{tabular}

    \vspace{0.2cm}
    \begin{tabular}{|c|c|c|c|c|c|}
        \hline
         & DD2\_M135-135 & DD2\_M180-108 & BLh\_M1146-1635 & SFHo\_M135-135 & SLy\_M145-125 \\
        \hline
        DD2\_M135-135 & $\dots$  & $322.4$ & $15.5$ & $45.0$ & $166.2$ \\
        \hline
        DD2\_M180-108 & $-322.4$ & $\dots$  & $-306.9$ & $-277.4$ & $-156.2$ \\
        \hline
        BLh\_M1146-1635 & $-15.5$ & $306.9$ & $\dots$  & $29.5$ & $150.7$ \\
        \hline
        SFHo\_M135-135 & $-45.0$ & $277.4$ & $-29.5$ & $\dots$  & $121.2$ \\
        \hline
        SLy\_M145-125 & $-166.2$ & $156.2$ & $-150.7$ & $-121.2$ & $\dots$  \\
        \hline
    \end{tabular}

    \vspace{0.2cm}
       \begin{tabular}{|c|c|c|c|c|}
        \hline
         DD2\_M135-135 & DD2\_M180-108 & BLh\_M1146-1635 & SFHo\_M135-135 & SLy\_M145-125 \\
        \hline
        $5.9$ & $301.2$ & $22.1$ & $53.7$ & $174.6$ \\
        \hline

    \end{tabular} 

        \caption{Same as in Table \ref{tab:bayes_1} but following the approach described in Sec. \ref{subsec:bayesian_II}.}

    \label{tab:bayes_2}

\end{table*}

In Table \ref{tab:bayes_1}, we show for different simulations (from left to right), the logarithm of the Bayes factors comparing our models as described in Sec.~\ref{subsec:bayesian_I}.
The top rows compare the ejecta profiles with an extended tail among the five different simulations, whereas the middle panels compare the ejecta profiles with a sharp cutoff.
	The analysis for the extended ejecta tail scenario only marginally favors DD2\_M135-135 over the other simulations we considered.
	On the contrary, ejecta profiles with a sharp cutoff present much larger differences.
	Finally, in the bottom row, we show the logarithm of the Bayes factors comparing the two approaches to model the ejecta profile (with an extended tail and with a sharp cutoff) for a given NR simulation.
 This demonstrates the larger likelihood of the extended tail scenarios by a significant margin.

In Table \ref{tab:bayes_2}, we show the same Bayes factors but following the approach described in Sec.~\ref{subsec:bayesian_II}.
	In this case, the strong constraint that the observed time delay for the arrival of $\gamma$ rays imposes on the time delay for jet launching time largely modifies the inference. 
	The two scenarios with a BH engine are favored, showing Bayes factors greater than one when compared with the cases with a magnetar engine. While the two simulations with a BH engine are comparable with each other, SFHo\_M135-135 turns out to be marginally better. 	
	In case of a sharp cutoff for the ejecta profile, DD2\_M135-135 is the most favored whereas DD2\_M180-108 is the worst by a significant margin, as can also be inferred from the results presented in Fig.~\ref{fig:parameter_estimation}.
	The comparison between these two modeling approaches for each simulation shows a clear preference toward the extended tail scenarios by a large margin.

\end{document}